\documentclass[aps,superscriptaddress]{revtex4}
\usepackage[T1]{fontenc}
\usepackage[utf8]{inputenc}
\usepackage{mathtools}
\usepackage{graphicx}
\usepackage{caption}
\usepackage{subcaption}
\usepackage{xcolor}
\usepackage{soul}
\usepackage{amsmath,amssymb}
\usepackage[colorlinks=true, pdfstartview=FitV, linkcolor=blue, citecolor=blue, urlcolor=magenta, breaklinks=true]{hyperref}
\usepackage{orcidlink}

\begin{document}

\title{An Undergraduate Approach to the Quantum Hadrodynamics and the Physics of Neutron Stars}

\author{Luiz L. Lopes}
\email{llopes@cefetmg.br}

\affiliation{Centro Federal de Educação Tecnológica de Minas Gerais Campus VIII, \\ CEP 37.022-560, Varginha, MG, Brazil}

\begin{abstract}

In this tutorial, I discuss how to model a neutron star from the Quantum Hadrodynamics microscopic approach. After a brief discussion about hydrostatic equilibrium, I discuss the role of each meson of the model and how to calculate the corresponding equation of state and the expected values. Each meson is introduced individually. Its effects are analyzed from both an analytical and a numerical point of view. To explicitly show the effects of a given meson, the coupling constant is varied in an arbitrary range before being fixed to reproduce well-known constraints. This work is intended for late undergraduate students as well as early graduate students. The equation of states is obtained from the statistical mechanics formalism, which is more familiar to students at this stage of their research career, instead of the traditional quantum field theory formalism.

\end{abstract}

\maketitle

\section{Introduction} \label{introduction}

Neutron stars are compact objects formed after the lives of massive ordinary stars. A low-mass star ($M~<~8M_\odot$) usually ends its life as a planetary nebula, which leads to the formation of a white dwarf. White dwarfs are compact objects whose gravitational stability comes from the electron degeneracy.
Massive stars end their lives in a supernova explosion. For a supernova remnant below approximately 3 $M_\odot,$ the increasing density induces electron capture on protons and prevents neutron $\beta$-decay, causing the so-called neutronization of matter~\cite{pizzochero2010}, hence the name neutron star. As neutrons are unstable, a small fraction of them is expected to decay into protons and electrons to maintain chemical equilibrium.

Historically, the theory of neutron stars can be traced back to the work of Lev Landau in the early 1930s, even before the discovery of the neutron~\cite{Landau1932,Yakovlev:2013}. In 1939, Oppenheimer and Volkoff derived the relativistic equation of hydrostatic equilibrium and applied it to pure neutron matter equilibrated only by neutron degeneracy. They found a maximum mass around 0.7 $M_\odot$~\cite{TOV}. The discovery of the first pulsars by Jocelyn Bell and Antony Hewish~\cite{Bell1967} confirmed that degenerate hadronic matter exists in nature. Modern measurements reveal that neutron stars can reach masses as high as twice the mass of our Sun~\cite{Miller2021,Riley2021}. As neutron degeneracy pressure alone is not enough to produce such a high mass, this implies the existence of a strong repulsive force between the nucleons at low distances.

One way to model the interaction between the nucleons is the Quantum Hadrodynamics (QHD)~\cite{Serot_1992}. The QHD considers the byarons as the fundamental degree of freedom, while the interaction is mediated by the exchange of massive mesons. In this work, I discuss the mean field approximation of the QHD, its formalism, and how to obtain a realistic equation of state (EOS) to describe neutron stars. There are excellent texts about QHD~\cite{Serot_1992,Glenbook,debora-universe}, but they use a formalism beyond what a typical late undergraduate student is used to. In the same sense, there are excellent introductory texts about neutron stars~\cite{Silbar2004,Sagert_2006}, but they do not cover the QHD.

The primary goal of this tutorial is to fill that gap. I begin with a concise review of the conditions for hydrostatic equilibrium, both in the Newtonian and relativistic frameworks, leading to the Oppenheimer–Volkoff (OV) equations. The following sections are dedicated to constructing the equation of state (EOS), starting from a model of free neutron matter. I then introduce the QHD framework incrementally—one meson at a time—so the impact of each interaction becomes clear. The same step-by-step approach is applied to the inclusion of nonlinear meson self-interactions. All calculations are presented in as much detail as possible, using methods grounded in statistical mechanics, making the formalism accessible to students without requiring a background in quantum field theory.

\section{Hydrostatic Equilibrium}

In a star's interior, two forces act on a fluid: gravitation and pressure. Despite nuclear interaction, neutron stars are bound by gravity. Our first investigation focuses on studying the relationship between pressure and gravity to achieve hydrostatic equilibrium.

\subsection{Newtonian Case}

Consider a perfect fluid in hydrostatic equilibrium. In this case, its behavior obeys Euler's equation~\cite{LandauFluid}:

\begin{equation}
  \nabla p = {\bf f}, \label{euler}  
\end{equation}
where $p$ is the pressure and ${\bf f}$ is the force density. For a self-gravitating fluid, we have the traditional ${\bf f} = \rho {\bf g}$, where $\bf g$ is the gravitational field and $\rho = \rho(r)$ is the mass density. The gravitational field can be expressed as a gradient of the gravitational potential, ${\bf g} = - \nabla \Phi$, which, in turn, obeys Poisson's equation:

\begin{equation}
  \nabla^2 \Phi = 4\pi G \rho(r), \label{poisson}  
\end{equation}

 Eq.~\ref{euler} can be rewritten as:

\begin{equation}
  \frac{\nabla p}{\rho(r)} = - \nabla \Phi, \label{eq3}  
\end{equation}

Using the divergent on both sides in Eq.~\ref{eq3} and combining it with Eq.~\ref{poisson}:

\begin{equation}
 \nabla \cdot \bigg (\frac{\nabla p}{\rho(r)} \bigg ) =  -\nabla^2 \Phi  = - 4\pi G \rho(r) .
\end{equation}

Now we integrate over the volume of a sphere and apply the Gauss theorem to obtain the hydrostatic equilibrium equation in its Newtonian form:

\begin{eqnarray}
\frac{dp}{dr} = -\frac{GM(r)\rho(r)}{r^2}, 
\end{eqnarray}
where we define

\begin{equation}
 M~\equiv~ \int 4\pi  r^2 \rho(r) dr .  
\end{equation}

The initial conditions are M(0) = 0 and $p(0) = p_0$. Moreover, when the pressure goes to zero, we claim that the star's surface was reached, i.e., $p(R) = 0$~\cite{Sagert_2006}. \\

\subsection{Full relativistic equilibrium equations}

The relativistic hydrostatic equilibrium equations are derived directly from Einstein's field equations:

\begin{equation}
 R_{\mu\nu} - \frac{1}{2}Rg_{\mu\nu} = -\frac{8\pi G}{c^4}T_{\mu\nu}  , \label{einsteinFE}
\end{equation}
where $R_{\mu\nu}$ is the Ricci curvature tensor, $R = R_\mu^{~\mu}$ is the scalar curvature, $g_{\mu\nu}$ is the metric tensor \footnote{With a metric signature (+,-,-,-).}, and $T_{\mu\nu}$ is the energy-momentum tensor. All the nomenclature and definitions used in this section are the same as presented in the book of Foster and Nightingale~\cite{Nightingale}~\footnote{Maybe more standard textbooks on general relativity are refs.~\cite{Carroll,Schutz}.}. 

The Ricci tensor is given as a function of the metric connections, $\Gamma^{\alpha}_{\mu\nu}$, and its derivative~\footnote{$\Gamma^{\alpha}_{\mu\nu}$ are also called Christoffel symbols}:

\begin{equation}
 R_{\mu\nu}~\equiv~\Gamma^\alpha_{\mu\alpha,~\nu}  - \Gamma^\alpha_{\mu\mu,~\alpha} + \Gamma^{\sigma}_{\mu\alpha}\Gamma^{\alpha}_{\sigma\nu} - \Gamma^{\sigma}_{\mu\nu}\Gamma^{\alpha}_{\sigma\alpha}.
\end{equation}

The metric connections are related to the derivatives of the metric tensor:

\begin{eqnarray}
  \Gamma^{\alpha}_{\mu\nu}~\equiv\frac{1}{2}~g^{\alpha\sigma}(g_{\sigma\nu,~\mu}  +g_{\mu\sigma,~\nu} - g_{\mu\nu,~\sigma}).
\end{eqnarray}

For the right side of Einstein's equation, we consider the energy-momentum tensor of a perfect fluid:

\begin{equation}
 T_{\mu\nu}   = (\rho + p/c^2)u_\mu u_\nu -p g_{\mu\nu}, \label{perfect}
\end{equation}
where $u_\mu$ is the four-velocity. By imposing hydrostatic equilibrium, we obtain $u_0 =c$ and $u_\mu$ = 0 for $\mu \neq 0$. In the same sense, imposing spherical symmetry, we have $T_{\mu\nu} = 0$ for $\mu \neq \nu$. Now, the metric for a spherically symmetric spacetime reads:

\begin{equation}
  g_{\mu\nu}dx^\mu dx^\nu = ds^2 = e^\Phi c^2dt^2 - e^\lambda dr^2 - r^2(d\theta^2 + \sin^2\theta d\phi^2) .\label{metric}
\end{equation}

Using Eq.~\ref{metric} to solve Einstein's field equations, we have:

\begin{equation}
 e^{-\lambda} \bigg (-\frac{\lambda'}{r} + \frac{1}{r^2} \bigg )   -\frac{1}{r^2} = -\frac{8\pi G}{c^4}\rho c^2, \label{elambda}
\end{equation}

\begin{equation}
 e^{-\lambda} \bigg (\frac{\Phi'}{r} + \frac{1}{r^2} \bigg )   -\frac{1}{r^2} = \frac{8\pi G}{c^4}p \label{ephi}.
\end{equation}

A third equation is obtained by taking the four-divergence of the energy-momentum tensor and setting it equal to zero\footnote{Setting this four-divergence equal to zero implies that gravity is not seen as a force in general relativity. The Newtonian counterpart is the Euler equation, which explicitly deals with the force density ${\bf f}$.}

\begin{equation}
 T^{\mu\nu}_{~~~~;\mu}   = 0 , \label{e4d}
\end{equation}
which give us

\begin{equation}
 \frac{dp}{dr} = - \frac{\Phi'}{2}(\rho c^2 + p). \label{epressure}
\end{equation}

Now, we can facilitate our calculation by imposing that the Schwarzschild metric is restored at the surface of a star, which gives us:

\begin{equation}
 e^{\lambda} =  \bigg (1 - \frac{2GM(r)}{c^2r} \bigg ), \quad \mbox{implying,} \label{lambdaS}   
\end{equation}

\begin{equation}
 \lambda' =     \bigg (1 - \frac{2GM(r)}{c^2r} \bigg ) \bigg [\frac{2G}{c^2r}\bigg (\frac{dM(r)}{dr} - \frac{M(r)}{r} \bigg ) \bigg  ] .\label{lambdaL}
\end{equation}

Now, by combining Eq.~\ref{ephi} with Eq.~\ref{lambdaS}, we obtain an expression for $\Phi'$:

\begin{equation}
 \Phi' = \bigg (\frac{8\pi G}{c^4}pr + \frac{2G M(r)}{c^2r^2} \bigg )\bigg ( 1 -\frac{2GM(r)}{c^2r} \bigg )^{-1} .  \label{phiL} 
\end{equation}

Finally, combining Eqs.~\ref{lambdaS} and~\ref{lambdaL} with Eq.~\ref{elambda}, and Eq.~\ref{epressure} with Eq.~\ref{phiL}, and we obtain the famous Oppenheimer-Volkoff (OV) equations~\footnote{Oppenheimer-Volkoff equations (OV) are sometimes called Tolman-Oppenheimer-Volkoff equations (TOV). However, as discussed in ref.~\cite{SEMIZ2016} the former is preferred.}:

\begin{equation}
M = \int 4\pi r^2 \rho(r) dr , \label{massOV}
\end{equation}
\begin{equation}
\frac{dp}{dr} = - \frac{GM(r)\rho(r)}{r}\bigg [1 + \frac{p}{\rho c^2}\bigg]\bigg [1 +\frac{4\pi p r^3}{M(r)c^2} \bigg ]\bigg [1 -\frac{2GM(r)}{c^2r} \bigg ]^{-1}. \label{OV}
\end{equation}

We can notice that the gravitational mass enclosed in radius R is the same in the Newtonian as well as in the relativistic formulation of gravity. However, the Newtonian hydrostatic equilibrium equation is only an approximation, valid for fluids with $p <<\rho c^2$, as well as bodies with a low mass/radius relation.
We see that each term in brackets is positive and is larger than one. Therefore, gravity in the relativistic formalism is stronger than its Newtonian counterpart.
In the Newtonian approximation, the mass density is the only source of the gravitational field and must be balanced by the pressure gradient.
In a full relativistic approach, besides the mass density, the pressure itself acts as a source of the gravitational field, present in the first two brackets. The last bracket points out that the curvature of spacetime also acts as an additional source of the gravitational field. For a relativistic fluid, the mass density must be replaced by the energy density: $\epsilon(r) \approx  \rho(r) c^2$.

The same initial conditions present in the Newtonian formalism are also present in the relativistic one ($p(0) = p_0,~M(0) =0,~p(R)=0$). However, one can notice that we have two equations, but three variables, $M(r),~p(r)~\mbox{and}~\epsilon(r)$. Another equation is necessary to obtain a solution of the OV equations. This is the role of nuclear physics: to obtain an equation of state (EOS), i.e., pressure as a function of the energy density, $p = p(\epsilon)$. With the EOS, we can obtain a whole family of solutions for different values of central pressures. Constructing realistic EOSs is the main goal of the next chapters. \\

\section{Free neutron matter}

Here we start our path to construct EOS and solve the OV equations. From this point on, we use natural units: $\hbar = c = k_B = 1$.

\subsection{Statistical Physics}

Consider that some particles (in our case, neutrons) are confined in a region of space, called the $\mu$-space, which is represented by a cloud of N points in a 6D space (3 for momentum and 3 for position~\cite{Huang2001}). We can divide the $\mu$-space into 6D cells, subject to the condition that each cell is large enough to contain a large number of particles, and yet small enough to be considered infinitesimal on a macroscopic scale. Moreover, each particle has $\gamma$-fold degeneracy~\footnote{$\gamma$-fold degeneracy means the number of identical particles that can present the same energy eigenvalue.}, and the minimum cell size according to the uncertainty principle is $h^3$; we therefore read~\cite{Huang2001}:

\begin{equation}
d \tau = \gamma \frac{d^3kd^3r}{(2\pi^3)} \label{cell} ,
\end{equation}
where $k$ represents the momentum.  The number of particles in a given cell is called the occupation number $dn$, and the distribution function, $f(k,r)$, is the occupation number per unit of (6D) volume:

\begin{equation}
  dn = f(k,r)d\tau  . \label{dtau}
\end{equation}

The total energy and number of particles are therefore:

\begin{eqnarray}
   E =  \frac{\gamma}{(2\pi)^3}\int E(k,r)f(k,r) d^3kd^3r, \nonumber \\
   N =   \frac{\gamma}{(2\pi)^3} \int f(k,r)d^3kd^3r , \label{totais}
\end{eqnarray}
where $E(k,r)$ is the energy eigenvalue of a particle of momentum $k$ in the position $r$. To solve the OV equations, we must obtain their density values, ie, the energy density ($\epsilon =E/V$) and the number density ($n =N/V$). In this case, we have $f(k,r) = f(k)$ and

\begin{eqnarray}
  \epsilon =  \frac{\gamma}{(2\pi)^3} \int E(k)f(k) d^3k, \nonumber \\
   n =  \frac{\gamma}{(2\pi)^3}   \int f(k)d^3k     . \label{end}
\end{eqnarray}

Neutrons are fermions with spin-1/2, therefore they have a degeneracy $\gamma = (2S +1) =2$ and obey the Fermi-Dirac distribution~\cite{Sagert_2006,Huang2001}:

\begin{equation}
f(k) = f(E) =  \frac{1}{\exp[(E - \mu)/T] +1} , \label{FD}    
\end{equation}
where $E$ is the energy eigenvalue, $\mu$ is the chemical potential, and $T$ is the temperature. Due to the high densities reached in neutron stars' interiors, the Fermi energy of neutrons can reach values above 1 GeV. On the other hand, the temperature of an ordinary neutron star is below 0.1 MeV. Due to this, the effects of temperature can be neglected, and $T = 0$ K is a good approximation~\footnote{see also Sec. 9.3 of ref.~\cite{Huang2001}.}
In this case, $\mu = E_F$ and the Fermi-Dirac distribution becomes the Heaviside step function:

\begin{eqnarray}
  f(E) = \left\{ \begin{array}{ll}
         1 & \mbox{if $E \leq E_F$};\\
        0 & \mbox{if $E > E_F$}.\end{array} \right.  \label{heaviside}
\end{eqnarray}

This means that all states with the energy below the Fermi energy ($E_F$) are occupied, and all those above are empty. In the momentum space, the occupied states lie within the Fermi sphere of radius $k_F$, so-called Fermi momentum.
Finally, the energy eigenvalue of neutrons is given by Einstein's relation:

\begin{equation}
 E =\sqrt{M^2 +k^2} , \label{freeNE}   
\end{equation}
where the neutron mass $M~\approx939$ MeV. The chemical potential is defined at the Fermi momentum:

\begin{equation}
  \mu = E_F = \sqrt{M^2 +k_F^2}
\end{equation}
The neutron number and energy density are therefore:

\begin{eqnarray}
  n = \frac{2}{(2\pi)^3}\int_0^{k_F}d^3k = \frac{8\pi}{(2\pi)^3}\int_0^{k_f}k^2 dk = \frac{k_F^3}{3\pi^2}  , \label{numberd} \\
  \epsilon = \frac{2}{(2\pi)^3}\int_0^{k_F}Ed^3k = \frac{1}{\pi^2}\int_0^{k_f}\sqrt{M^2 +k^2}k^2dk . \label{energydensity}
\end{eqnarray}

\subsection{Dirac Equation and Dirac Lagrangian}

Let us take a step back and ask: what is the relativistic equation for fermions that gives us the correct energy eigenvalue presented in Eq.~\ref{freeNE}?

The answer is the so-called Dirac Equation~\cite{griffiths_part,GreinerRQM}:

\begin{equation}
  (i\gamma^\mu\partial_\mu - M)\psi = 0 , \label{dirac}  
\end{equation}
where $\psi$ is the Dirac Fermi field, a four-element spinor, $M$ is the mass of the fermion, and $\gamma^\mu$ are the so-called Dirac matrices, a set of 4x4 matrices that can be expressed as a function of the traditional Pauli matrices:

\begin{eqnarray}
   \gamma^0 = \left( \begin{array}{ll}
         1 & 0\\
        0 & -1 \end{array} \right ) , \quad 
        \gamma^i = \left( \begin{array}{ll}
         0 & \sigma^i\\
        -\sigma^i & 0 \end{array} \right ) .\label{DMatrix}
\end{eqnarray}

Let us take another step back and ask: what is the Lagrangian (or more rigorously, the Lagrangian density) that produces the Dirac equation? 

The answer is the Dirac Lagrangian:

\begin{equation}
 \mathcal{L} = \bar{\psi}[i\gamma^\mu\partial_\mu -  M]\psi   ,\label{diraclag}
\end{equation}
now applying the Euler-Lagrange (E-L) equations:

\begin{eqnarray}
 \partial_\alpha \bigg (\frac{\partial \mathcal{L}}{\partial(\partial_\alpha q)} - \frac{\partial\mathcal{L}}{\partial q} \bigg ) = 0, \label{EL}   
\end{eqnarray}
considering $q = \bar{\psi}$, the Dirac equation (Eq.~\ref{dirac}) is obtained.

Obtaining the Dirac Lagrangian when we already have the Dirac equation may sound redundant, but when interactions are introduced, this step becomes crucial, once the Lagrangian is the starting point.

We are now going to solve the Dirac equation. Our main goal is to obtain the energy eigenvalue. The four-spinor can be written as:

\begin{eqnarray}
   \psi = \left( \begin{array}{l}
         u_1 \\
        u_2 \\
        u_3 \\
        u_4 \end{array} \right )   =     \left( \begin{array}{l}
         u_A \\
         u_B \end{array} \right ) , \quad \mbox{with} \quad 
         u_{A(B)} = \left( \begin{array}{l}
         u_{1(3)} \\
        u_{2(4)} 
        \end{array} \right ) .
\end{eqnarray}

Applying the Quantization Rules:

\begin{equation}
  k^0 = E = i\partial^0, \quad k^i = -i\partial^i, \label{quantrules}
\end{equation}
the Dirac Equation reads:

\begin{eqnarray}
    \bigg [\left( \begin{array}{ll}
         1 & 0\\
        0 & -1 \end{array} \right ) \cdot E~-~
         \left( \begin{array}{ll}
         0 & \vec{\sigma}\\
        -\vec{\sigma} & 0 \end{array} \right )\cdot \vec{k}~-~
        \left( \begin{array}{ll}
         1 & 0\\
        0 & 1 \end{array} \right )\cdot M \bigg]   \left( \begin{array}{l}
         u_A \\
         u_B \end{array} \right ) =0 ,
        \label{DS1}
\end{eqnarray}
rearranging,

\begin{eqnarray}
    \bigg [\left( \begin{array}{ll}
         (E -M) & - \vec{\sigma}\cdot \vec{k} \\
         \vec{\sigma}\cdot \vec{k} & -(E +M) \end{array} \right ) 
        \bigg ]   \left( \begin{array}{l}
         u_A \\
         u_B \end{array} \right ) =0 ,
        \label{DS2}
\end{eqnarray}
which is an algebraic equation for $E$. Solving, we obtain $E = \sqrt{M^2 +k^2}$, which at the Fermi momentum, $k =k_F$ and $T = 0$ K is also the chemical potential~\footnote{Actually, the complete solution is $E = \pm~\sqrt{M^2 +k^2}$. The negative sign is related to the antiparticles, which at $T = 0K$ play no role. See ref.~\cite{griffiths_part} for additional discussion. }. Although the result was expected, this shows the coherence of the Lagrangian formalism. \\

\subsection{Equation of State}

We calculated the number density $n$, and the energy density $\epsilon$. To solve the OV equations, we now need the pressure $p$.  It can be trivially obtained from (another) Euler equation~\footnote{Eq.~\ref{euler} is also derived from an Euler equation, as well as Eq.~\ref{EL} shares the name of Euler}~\cite{Blundell_CTP}:

\begin{equation}
E = TS -PV + \mu N  .
\end{equation}

The Euler equation per unit of volume and at $T = 0$ K can be rewritten as:

\begin{equation}
p = \mu n - \epsilon  . \label{pressure1}
\end{equation}

This equation correctly gives us the pressure for any fluid. Nevertheless, it is possible to obtain the pressure as a function of $k$ explicitly, directly from the thermodynamics:

\begin{equation}
  p = - \frac{\partial E}{\partial V} .  
\end{equation}
Considering $E = \epsilon V$ and $V = N/n$, with a fixed $N$ we obtain:

\begin{equation}
p = n^2 \frac{\partial (\epsilon/n)}{\partial n} , \label{thermopressure}
\end{equation}
resulting in:

\begin{equation}
  p =   \frac{1}{3\pi^2}\int_0^{kf} \frac{k^4 dk}{\sqrt{M^2 +k^2}} . \label{pressureFermi}
\end{equation}

\begin{figure*}[ht]
\begin{tabular}{ccc}
\centering 
\includegraphics[scale=.58, angle=270]{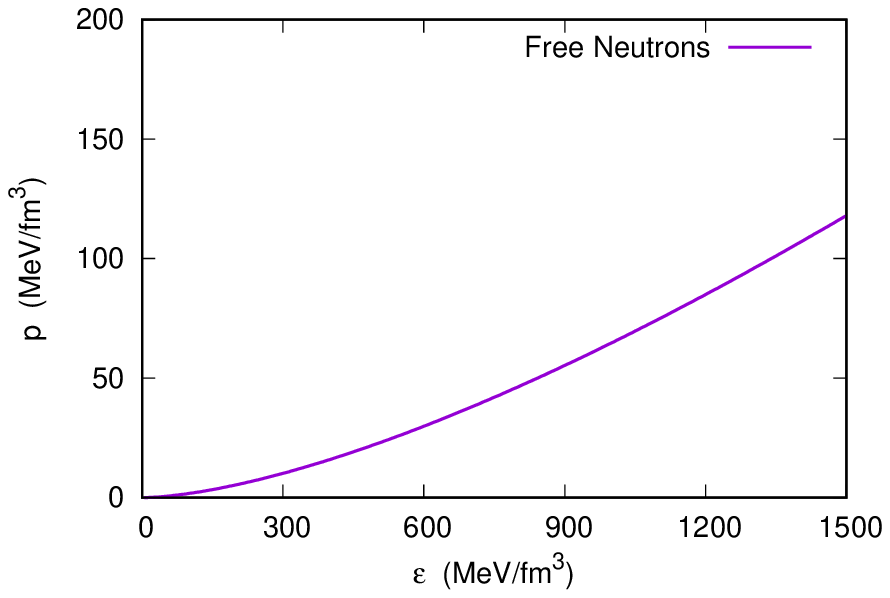} &
\includegraphics[scale=.58, angle=270]{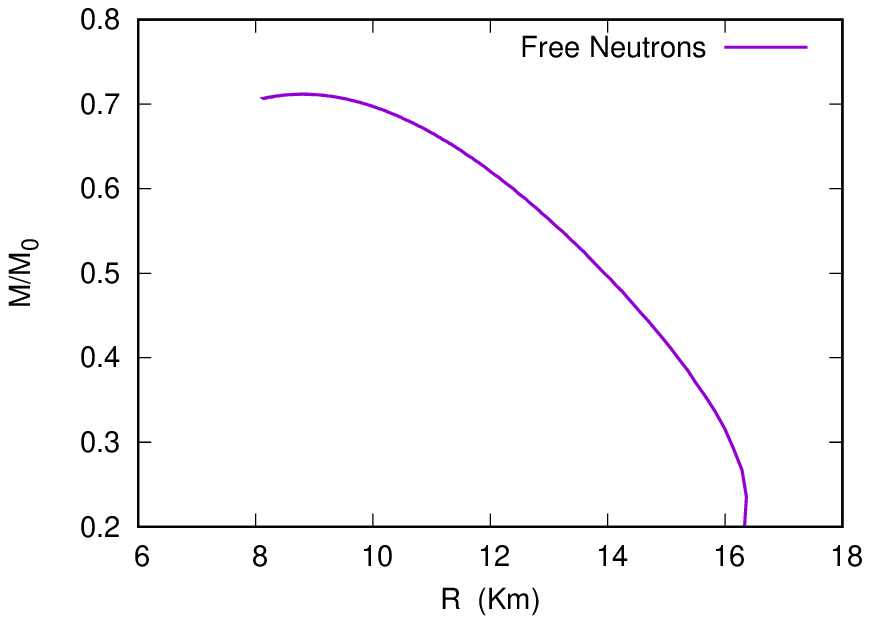} \\
\end{tabular}
\caption{\textbf{a}) EOS and (\textbf{b}) mass-radius relation for free neutron matter.\label{F1}} 
\end{figure*}

Once the EOS is obtained, we can use it as an input to the OV equations.
The EOS and the OV solution for free neutron gas are displayed in Fig.~\ref{F1}.

This result is the same as that obtained by Oppenheimer and Volkoff in the late 1930s~\cite{TOV}. Each dot in curve {\bf (b)} represents a different neutron star, with a different initial condition $p(0) =p_0$. As we increase the value of $p_0$, the new neutron star has a larger mass, up to a maximum value. For free neutrons, the maximum mass is only 0.71 $M_\odot$. It is clear that free neutron matter cannot represent the true state of dense nuclear matter. 
Today, the most massive well-measured pulsar is the  PSR J0740+6620, with a mass of 2.08 $\pm$ 0.07 $M_\odot$ and a radius in the range of 11.41 km $<~R~<$ 13.69 km~\cite{Riley2021,Miller2021}. Any realistic EOS must be able to fulfill this constraint. \\

\section{The scalar attractive force: the $\sigma$ meson}

Until the early 1930s, before the discovery of neutrons, it was believed that nuclei were a composite of protons and electrons~\cite{Lilley2001}. This was justified by the fact that the atomic nuclei have a mass approximately a multiple of the mass of the proton. Take the He nuclei, for example. It was believed that it contains four protons and two electrons, which correctly describes its mass and charge. Moreover, the electrons in the nuclei were responsible for balancing the repulsion between protons. However, as soon as the neutron was discovered, it became clear that the atomic nucleus was composed of protons and neutrons.

A new question was then raised. What was responsible for keeping the nuclei stable? One answer was given by H. Yukawa in 1935, who proposed that the interactions between protons and neutrons are mediated by the exchange of a massive boson~\cite{Yukawa1935}. 
This nuclear potential must be strong enough to overcome the electrostatic repulsion between the protons, yet its range must be very limited. Yukawa proposed the following potential:

\begin{equation}
  V(r)  = -\frac{g^2}{4\pi}\frac{e^{-mr}}{r} , \label{yuk1}
\end{equation}
where $m$ is the mass of (at the time) a hypothetical boson.
The Yukawa potential leads to the discovery of the pion, which is the main responsible for the nuclear attraction at low energies. Therefore, it is natural to think that our effective model should include pions.

However, from a modern point of view, pions are not indicated in the QHD. The first reason is that pions are pseudo-scalar mesons, and pseudo-scalar mesons do not contribute in the mean-field approximation (MFA)~\cite{debora-universe}. 
Secondly, pions are only weakly attractive. Although one pion exchange gives us a good description at large distances (here, large means 1.5 fm $<~r~<$2.5 fm), it does not correctly describe the stronger interaction at moderate distances (0.7 fm ~$<~r~<~1.5$~fm). To overcome this issue, several authors use a two-pion exchange~\cite{Pion1,Pion2}.

From an effective theory point of view, two-pion exchange is equivalent to a scalar-meson exchange~\cite{Barker1967,DURSO1980}. Here we consider the Yukawa coupling from Eq.~\ref{yuk1} for the scalar $\sigma$ meson. Unlike pions, the true nature of the $\sigma$ meson is not well understood, and it is based on the $f_0(500)$ resonance~\cite{PDG2020}. Here, we use a mass of 512 MeV.

The meson-nucleon interaction is introduced via minimal coupling~\cite{griffiths_part}. Its Lagrangian is called the Yukawa-Dirac Lagrangian and reads:

\begin{equation}
  \mathcal{L} = g_s(\bar{\psi}\sigma\psi)  \label{yukL1}
\end{equation}

Moreover, as the $\sigma$ meson is present, we also need the Lagrangian of free spin-0 particles:

\begin{equation}
\frac{1}{2} (\partial^\mu \sigma \partial_\mu \sigma   - m_s ^2\sigma^2) \label{sigmaf}
\end{equation}
where $m_s$ is the mass of the $\sigma$ meson. Combining Eq.~\ref{diraclag} with Eq.~\ref{yukL1} and Eq.~\ref{sigmaf} we have the Lagrangian of neutrons interacting via $\sigma$-mesons exchange.

\begin{equation}
 \mathcal{L} =  \bar{\psi}[i\gamma^u\partial_\mu - (M -g_s\sigma)]\psi + \frac{1}{2} (\partial^\mu \sigma \partial_\mu \sigma   - m_s^2 \sigma^2) .   \label{sigmaL}
\end{equation}

Applying the Euler-Lagrange equations to the $\sigma$ meson we obtain:

\begin{equation}
  (\square^2 + m_s^2)\sigma = g_s(\bar{\psi}\psi).  \label{KG}
\end{equation}
where $\square^2$ is the d'Alembertian.  Eq~\ref{KG} is called Klein-Gordon equation~\cite{griffiths_part}. In this case, with a source,  $g_s(\bar{\psi}\psi)$.

\subsection{Mean Field Approximation}

Exact solutions of Eq.~\ref{KG} are very complex, once neither the mesonic field nor the baryon-antibaryon field can be treated as point-like particles but rather as objects with intrinsic structure due to the implied (virtual) meson and baryon-antibaryon loops~\cite{Serot_1992}. A way to overcome this difficulty is via the mean-field approximation.

If the baryon number is large enough, the quantum fluctuation can be ignored, and we can replace the mesonic field $\sigma$ by its expected value:

\begin{equation}
  \sigma\to \langle \sigma \rangle \equiv \sigma_0,  
\end{equation}
in this regime, the mesonic field is stationary and independent of
space and time, which imply $\square^2 \sigma = 0$ (as well the so-called kinetic term vanishes: $\partial^\mu\sigma\partial_\sigma =0$). Eq.~\ref{KG} can be written as:

\begin{equation}
g_s\sigma_0 = \bigg (\frac{g_s}{m_s}\bigg )^2 \langle \bar{\psi}\psi \rangle . \label{sigma1}
\end{equation}

The quantity $\langle \bar{\psi}\psi \rangle$ is called scalar density. Its value is well known in the literature~\cite{Serot_1992,Glenbook,debora-universe}:

\begin{equation}
  \langle \bar{\psi}\psi \rangle = n^S = \frac{\gamma}{2\pi^2}\int_0^{k_f} \frac{M^{*}k^2 dk}{\sqrt{M^{*2} + k^2}} , \label{scalardensity}
  \end{equation}
where $\gamma$ is again the degeneracy number. The quantity $M^*$ is called nucleon effective mass, or Dirac mass. It is given by:

\begin{equation}
  M^* = M - g_s\sigma_0,  \label{effectivemass}
\end{equation}
implying that the Eq.~\ref{sigma1} must be solved self-consistently. In MFA, the mesonic fields are classical, and the nucleons behave like a free Fermi gas immersed in a background potential. Furthermore, the higher the density, the higher the accuracy of MFA.

Now, applying the E-L equations for the neutron Dirac field, we obtain:

\begin{equation}
 (i\gamma^\mu\partial_\mu - M^*)\psi = 0,   
\end{equation}
which is the same as Eq.~\ref{dirac} but with an effective mass $M^*$.
Therefore, we have:

\begin{equation}
 E = \sqrt{M^{*2} + k^2}. \label{energysigma}   
\end{equation}

Using the Fermi-Dirac statistic, we can obtain the energy density for the neutrons:

\begin{equation}
  \epsilon_N =   \frac{1}{\pi^2}\int_0^{k_f}\sqrt{M^{*2} +k^2} k^2 dk,
\end{equation}
as we also have the $\sigma_0$ field, we also must take its contribution to the total energy density.  The energy density of the mesonic fields is calculated in MFA from their expected values in the Hamiltonian: $\epsilon_M = \langle \mathcal{H} \rangle = - \langle \mathcal{L} \rangle$~\cite{Glenbook}:

\begin{equation}
 \epsilon_M = \frac{1}{2}m_s^2\sigma_0^2.   \label{edsigmaalone}
\end{equation}

The total energy density is therefore the sum of $\epsilon_N$ and $\epsilon_M$: 

\begin{equation}
 \epsilon = \frac{1}{\pi^2}\int_0^{k_f}\sqrt{M^{*2} +k^2} k^2 dk  + \frac{1}{2}m_s^2\sigma_0^2.   \label{edsigma}
\end{equation}

The same reasoning must be followed for the pressure. The neutron degeneracy pressure is calculated from Eq.~\ref{thermopressure}, and it is analogous to Eq.~\ref{pressureFermi} for an effective mass $M^*$:

\begin{equation}
 p_N = \frac{1}{3\pi^2}\int_0^{k_f} \frac{k^4 dk}{\sqrt{M^{*2} +k^2}},   
\end{equation}
while the pressure of the mesonic field in MFA is given by: $p = \langle \mathcal{L} \rangle$:

\begin{equation}
   p_M = - \frac{1}{2}m_s^2\sigma_0^2.   
\end{equation}

The total pressure is again the sum of the neutron and the mesonic contributions:

\begin{equation}
  p = \frac{1}{3\pi^2}\int_0^{k_f} \frac{k^4 dk}{\sqrt{M^{*2} +k^2}} - \frac{1}{2}m_s^2\sigma_0^2.  \label{psigma}
\end{equation}

Alternatively, Eq.~\ref{pressure1} can also be used. \\

\subsubsection{Stationarity of Energy Density}

Eq.~\ref{scalardensity} was not justified yet. We can nevertheless calculate the classical background field of the MFA by imposing the energy density to be stationary at fixed baryon density, $\partial \epsilon/\partial \sigma_0 = 0$:

\begin{equation}
 \bigg (\frac{\partial \epsilon}{\partial \sigma_0} \bigg ) = \frac{-g_s}{\pi^2} 
 \int_0^{k_f} \frac{M^{*} k^2 dk}{\sqrt{M^{*2} + k^2}} +m_s^2\sigma_0 = 0 .\label{sigmafield1}
\end{equation}

Rearranging this equation, Eq.~\ref{sigma1} and Eq.\ref{scalardensity} are recovered (with $\gamma =2$). \\

\subsection{Numerical Results}

\begin{figure*}[ht]
\begin{tabular}{ccc}
\centering 
\includegraphics[scale=.58, angle=270]{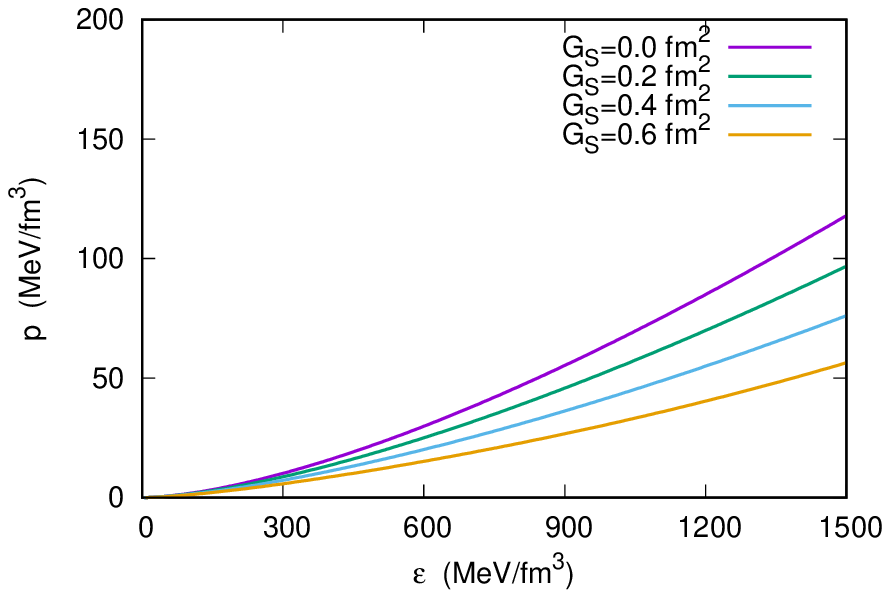} &
\includegraphics[scale=.58, angle=270]{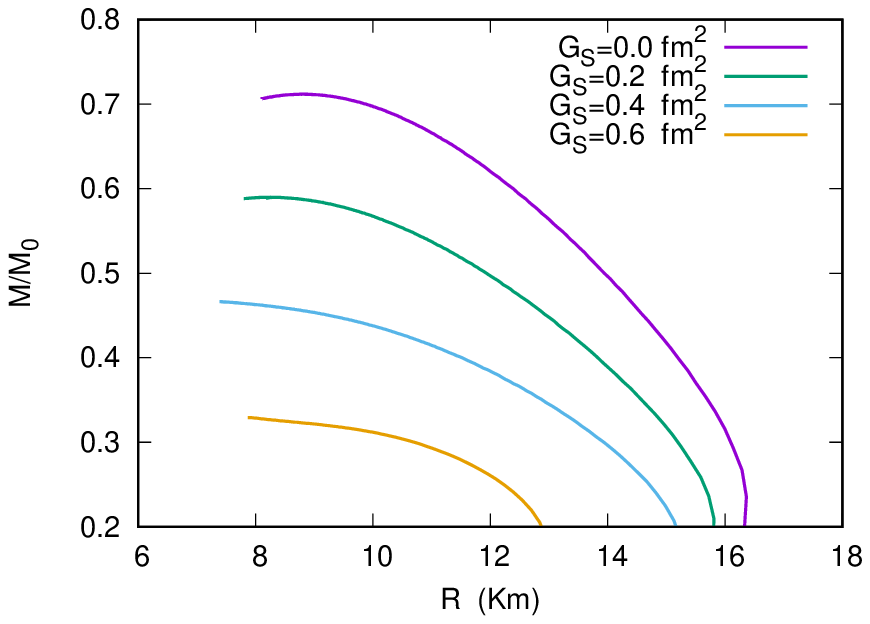} \\
\end{tabular}
\caption{(\textbf{a}) EOS and (\textbf{b}) mass-radius relation for different values of $G_S$. The softer the EOS, the lower the maximum mass. \label{F2}} 
\end{figure*}

Let us play with the numbers. We now analyze the effects of the $\sigma$ meson on the equation of state and the mass-radius relation from the OV solutions individually analyzing the four components, $\epsilon_N$, $\epsilon_M,~p_N$ and $p_M$.
 The main effect of the $\sigma$ field is to reduce the neutron mass. This implies a reduction of $\epsilon_N$, but an increase of $p_N$. However, as $\epsilon_N~>p_N$, the reduction in the energy density is more significant. The contribution of the mesonic field is the same for the energy density and the pressure, but with exchanged signals ($p_M = - \epsilon_M$). As $\epsilon_N~>~p_N$, the relative contribution of the mesonic field affects the pressure more than the energy density. Consequently, for a fixed Fermi momentum, both the energy density and the pressure are reduced, but the effects are stronger in the pressure.
The ratio $p/\epsilon$ is related to the so-called stiffness of the EOS.
A large value of $p/\epsilon$ gives us a stiff EOS~\footnote{The $p/\epsilon$ ratio must always be lower than one to keep causality. In QHD formalism, this condition is always satisfied.} A low value of $p/\epsilon$ give us a soft EOS. As
the value of $\sigma$ depends on the coupling constant $g_s$\footnote{More precisely, the $\sigma$ field depends on the ratio $(gs_/m_s)$, but $m_s$ is fixed.}, increasing it will soften the EOS. Let us make the following definition:

\begin{equation}
 G_S ~\equiv~ \bigg ( \frac{g_s}{m_s} \bigg )^2 ,  \label{GS} 
\end{equation}
which effectively gives us the strength of the $\sigma$ field. The EOS and the mass-radius relation for different values of $G_S$ are presented in Fig.~\ref{F2} \\

As can be seen, as we increase the value of the attractive field $\sigma$, we make the EOS softer, and consequently, the maximum mass is reduced.
In realistic EOSs, the value of $G_S$ is much larger because there are other mesons involved in the process. Here, we use small values of $G_S$ because lager values would produce unstable EOSs, where $dp/d\epsilon <0$~\footnote{Le Chatelier's principle, see ref.~\cite{Glenbook}.}.  \\

\section{ The repulsive $\omega$ meson and the $\sigma-\omega$ model.}

Effective nuclear models must ultimately describe, as accurately as possible, atomic nuclei. The fact that the nuclei are stable leads us to introduce the $\sigma$ meson. Another feature is that the density of several nuclei is approximately the same. This indicates that the nuclear force saturates, which implies that at a very short distance ($r <0.7$ fm) the nuclear force must become repulsive~\cite{Lilley2001}.

In QHD, the repulsive force is represented by a vector field. In this case, the $\omega$ meson. Unlike the scalar channel, which is based on the loose $f(500)$ resonance, the $\omega$ meson is a well-known vector meson with a mass of 783 MeV~\cite{PDG2020}.

A relativistic model containing both, the $\sigma$ and $\omega$ mesons is called the $\sigma-\omega$ model, and was introduced in many-body calculations by J. Walecka in 1974~\cite{WALECKA1974}. The effective potential now reads:

\begin{equation}
  V(r)  = \frac{g_\omega^2}{4\pi}\frac{e^{-m_\omega r}}{r} -\frac{g_s^2}{4\pi}\frac{e^{-m_sr}}{r} , \label{yuk2}
\end{equation}

From a relativistic point of view, the $\omega$ interaction is also introduced via minimum coupling with a Yukawa-Dirac Lagrangian. However, the $\omega$ meson is a spin-1 particle. Its Lagrangian is: 

\begin{equation}
 \mathcal{L} =  -g_\omega(\bar{\psi}\gamma^\mu\omega_\mu\psi) 
 - \frac{1}{4}\Omega^{\mu\nu}\Omega_{\mu\nu} + \frac{1}{2}m_\omega^2\omega^\mu\omega_\mu ,
\end{equation}
where $\Omega_{\mu\nu} = \partial_\mu\omega_\nu - \partial_\nu\omega_\mu$. The first term corresponds to the interaction of the nucleons via an exchange of the vector meson $\omega$, while the others represent the free Lagrangian of spin-1 particles, containing a kinetic and a mass term. The total Lagrangian for the $\sigma-\omega$ model reads~\cite{Serot_1992,Glenbook,debora-universe,WALECKA1974}:

\begin{equation}
 \mathcal{L} = \bar{\psi}[\gamma^\mu(i\partial_\mu -g_\omega\omega_\mu) - (M - g_s\sigma)]\psi +\frac{1}{2}(\partial^\mu\sigma\partial_\mu\sigma - m_s^2\sigma^2)    - \frac{1}{4}\Omega^{\mu\nu}\Omega_{\mu\nu} + \frac{1}{2}m_\omega^2\omega^\mu\omega_\mu . \label{sigma-omega-L}
\end{equation}

Now applying Euler-Lagrange equations with respect to the $\omega^\mu$ field, and considering that $\partial_\mu\partial^\nu\omega_\nu = 0$ due to the nucleon conservation (see ref.~\cite{Glenbook} for additional discussion), we obtain: 

\begin{equation}
 (\square^2 + m_\omega^2)\omega_\mu = g_\omega(\bar{\psi}\gamma_\mu\psi) , \label{klein2} 
\end{equation}
which is also a Klein-Gordon equation with a source $g_\omega(\bar{\psi}\gamma_\mu\psi)$.

Eq.~\ref{klein2} is also solved in MFA, with the same assumptions as in the $\sigma$ meson: the density is high enough that quantum fluctuations can be neglected, the field is stationary and independent of space and time, which allows us to replace $\omega_\mu$ field by its expected value.
Moreover, due to the space symmetry, all space components of the $\omega_\mu$ field vanish; only the temporal $\omega_0$ survives.

\begin{equation}
  \omega_\mu \to \delta^\nu_0\langle\omega_\nu\rangle = \omega_0.  
\end{equation}

Also, we have $\square^2\omega_0 = 0$, resulting in:

\begin{equation}
g_\omega\omega_0 = \bigg (\frac{g_\omega}{m_\omega}\bigg )^2 \langle \bar{\psi}\gamma^0\psi \rangle . \label{omega1}
\end{equation}

The quantity $\langle \bar{\psi}\gamma^0\psi\rangle$ is the traditional number density:

\begin{equation}
  \langle \bar{\psi}\gamma^0\psi \rangle =  n =   \frac{\gamma}{(2\pi)^3}\int d^3k = \gamma \frac{ k_F^3}{6\pi^2}. \label{numbd}
\end{equation}

Now, applying E-L in Eq.~\ref{sigma-omega-L} relative to the nucleon Dirac field, in MFA:

\begin{equation}
  [\gamma^0(i\partial_0 - g_\omega\omega_0) - i\gamma^j\partial_j  - M^{*}]\psi = 0]. \label{Dirac2}  
\end{equation}

Using the quantization rules from Eq.~\ref{quantrules}:

\begin{eqnarray}
    \bigg [\left( \begin{array}{ll}
         1 & 0\\
        0 & -1 \end{array} \right ) \cdot (E- g_\omega \omega_0)~-~
         \left( \begin{array}{ll}
         0 & \vec{\sigma}\\
        -\vec{\sigma} & 0 \end{array} \right )\cdot \vec{k}~-~
        \left( \begin{array}{ll}
         1 & 0\\
        0 & 1 \end{array} \right )\cdot M^{*} \bigg]   \left( \begin{array}{l}
         u_A \\
         u_B \end{array} \right ) =0 .
        \label{DSO1}
\end{eqnarray}

Eq.~\ref{DSO1} is analogous to Eq.~\ref{DS1} with a effective mass $M^*$ and an effective energy $E^* = E - g_\omega \omega_0$. Therefore its solution is $E^* = \sqrt{M^{*2} +k^2}$, which produces the energy eigenvalue of:

\begin{equation}
E = \mu = \sqrt{M^{*2} + k^2} + g_\omega\omega_0    .\label{ESO}
\end{equation}

Therefore, in MFA, the $\omega$ meson causes a shift in the energy, increasing its value.

 Once we obtain the energy eigenvalue of the neutron, its energy density can be obtained via the Fermi-Dirac distribution. On the other hand, the total energy density is obtained by summing the mesonic contribution of the Hamiltonian in MFA, $\langle \mathcal{H}\rangle  = - \langle \mathcal{L}\rangle$. We therefore have:

 \begin{equation}
  \epsilon = \frac{8\pi}{(2\pi)^3} \int_0^{k_f} [\sqrt{M^{*2} +k^2} + g_\omega\omega_0]  k^2 dk + \frac{1}{2}m_s^2\sigma_0^2 - \frac{1}{2}m_\omega^2\omega_0^2. \label{endso}
 \end{equation}

 The neutron pressure is obtained via Eq;~\ref{thermopressure}, and the mesonic contribution is $p = \langle \mathcal{L}\rangle$. The total pressure reads:

 \begin{equation}
  p = \frac{1}{3\pi^2}\int_0^{k_f}\frac{k^4dk}{\sqrt{M^{*2} +k^2}}   - \frac{1}{2}m_s^2\sigma_0^2 + \frac{1}{2}m_\omega^2\omega_0^2 ,\label{pso}
 \end{equation}
or alternatively, via Eq.~\ref{pressure1}.

The reader may notice that Eq.~\ref{numbd} has not yet been proved. This task can be accomplished by imposing the energy density to be stationary at fixed baryon density, $\partial \epsilon/\partial \omega_0  = 0$, as done for the $\sigma$ meson: 

\begin{equation}
\bigg ( \frac{\partial \epsilon}{\partial \omega_0} \bigg ) = \frac{1}{\pi^2}\int_0^{k_f}{g_\omega}k^2 dk -m_\omega^2\omega_0 = g_\omega n -m_\omega^2\omega_0 = 0.  \label{estaomgea}
\end{equation}

 Putting $\gamma = 2$, Eq~\ref{omega1} and Eq.~\ref{numbd} are recovered. Moreover, it is clear from  Eq.~\ref{estaomgea} that $(g_\omega\omega_0)\times n = m_\omega^2\omega_0^2$; which allows us to rewrite the energy density in a more standard way:

  \begin{equation}
  \epsilon = \frac{8\pi}{(2\pi)^3} \int_0^{k_f} [\sqrt{M^{*2} +k^2}]  k^2 dk + \frac{1}{2}m_s^2\sigma^2 + \frac{1}{2}m_\omega^2\omega_0^2. \label{endso2}
 \end{equation}

 Consequently,  as the mass term of the $\omega$ meson contributes equally to increasing both the energy density and the pressure,  the increase in pressure is much more significant due to the relative size of this term in the pressure when compared with its contribution to the energy density. Therefore, the $\omega$ meson stiffens the EOS. \\

 \subsection{Numerical results}

 We begin this section by defining the quantity $G_V$, which is analogous  to $G_S$ defined in Eq.~\ref{GS}:

 \begin{equation}
   G_V = \bigg (\frac{g_\omega}{m_\omega} \bigg )^2. \label{GV}  
 \end{equation}

\begin{figure*}[h!]
\begin{tabular}{ccc}
\centering 
\includegraphics[scale=.58, angle=270]{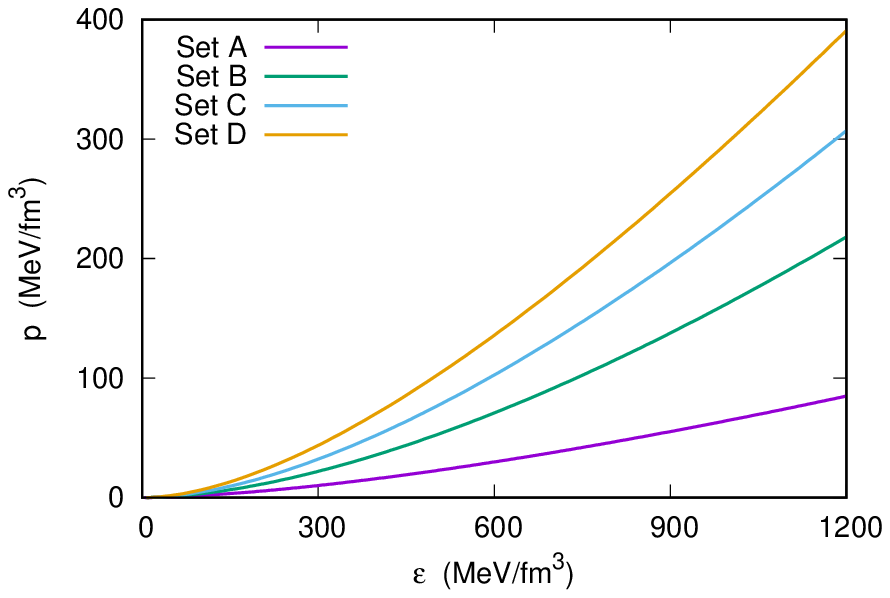} &
\includegraphics[scale=.58, angle=270]{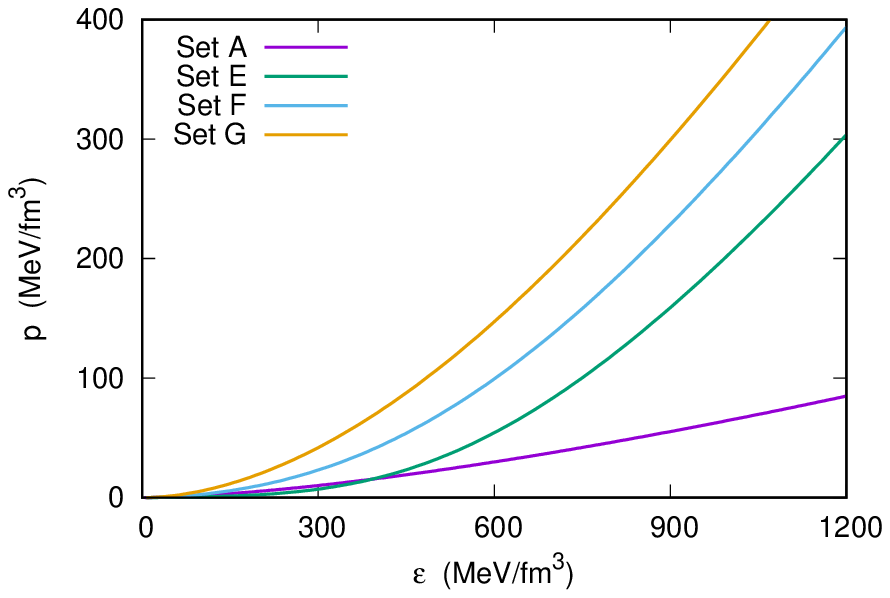} \\
\includegraphics[scale=.58, angle=270]{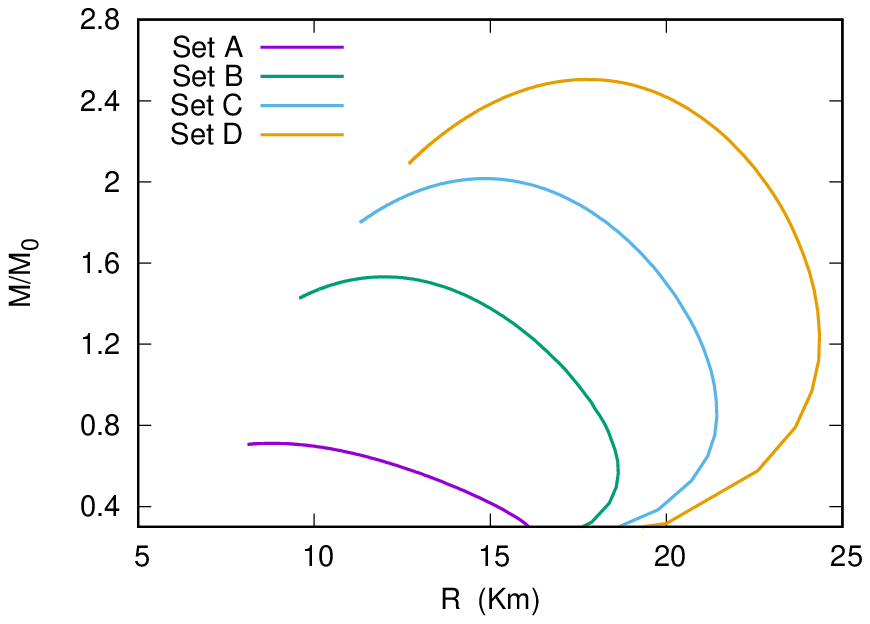} &
\includegraphics[scale=.58, angle=270]{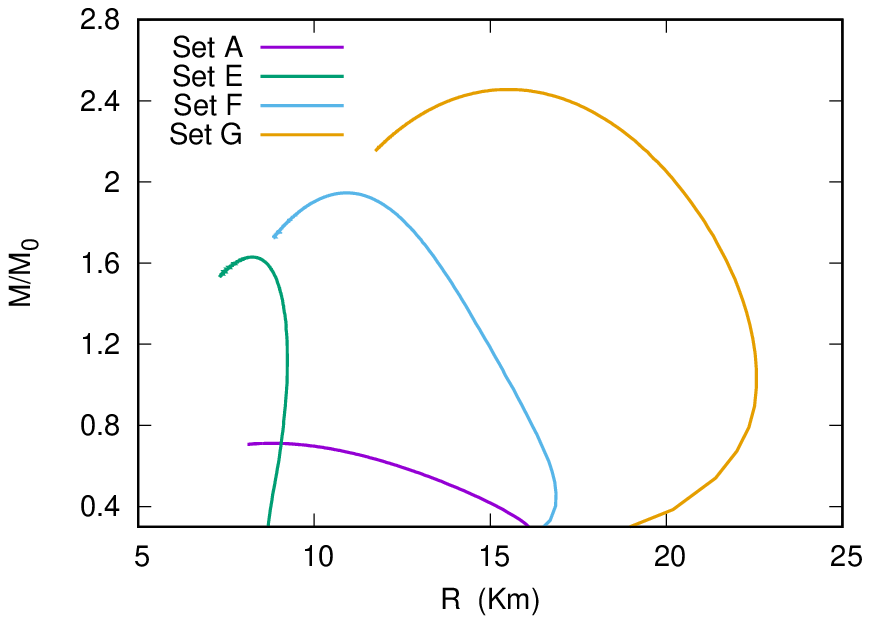} \\
\end{tabular}
\caption{EOSs and mass-radius relation for different values of $G_S$ and $G_V$ as follows: (\textbf{a}) EOSs for $G_S = 0$. (\textbf{b}) EOSs for $G_S~\neq 0$. (\textbf{c}) OV solution for  $G_S = 0$.  (\textbf{d})  OV solution for  $G_S~\neq~0$.\label{F3}}
\end{figure*}

We now show the effects of the $\omega$ meson in two different approaches. First, we take $G_S = 0$ to explicitly show the effects of the $\omega$ meson. Then, we fix the value of $G_S$ and vary $G_V$ to study their competition. The values utilized are presented in Tab.~\ref{tab1}. The corresponding EOSs and the mass-radius relation obtained by solving the OV equations are presented in Fig.~\ref{F3}, and the main results are summarized in Tab.~\ref{T2}.

\begin{center}
\begin{table}[h]
\begin{center}
\begin{tabular}{ccc}
\toprule
\textbf{Set}	& \textbf{$G_S$ (fm$^2$)}	& \textbf{$G_V$ (fm$^2$)}\\
Set A		& 0.00			& 0.00\\
Set B		& 0.00			& 1.50\\
Set C		& 0.00			& 3.00\\
Set D		& 0.00			& 5.00\\
Set E		& 4.50			& 3.00\\
Set F		& 4.50			& 5.00\\
Set G		& 4.50			& 8.00\\
\toprule
\end{tabular}
\caption{Different values of $G_S$ and $G_V$.\label{tab1}}
\end{center}
\end{table}
\end{center}

We can see that the main effect of the $\omega$ meson is to stiffen the EOSs and consequently to increase the maximum mass. The higher the value of $G_V$, the higher the maximum mass. From Set B to Set D we use $G_S = 0$, while from Set E to Set G we use $G_S \neq 0.$

We can notice that similar colors have similar maximum mass values. Explicitly, Set B with Set E has a maximum mass of around 1.6$M_\odot$, Set C with Set F around 2.0 $M_\odot$, and Set D with Set G around 2.5 $M_\odot$. Yet, they present different values for their radii, not only in the maximum masses but especially around the canonical 1.4$M_\odot$~\footnote{The name canonical star for 1.4$M_\odot$ comes from the fact that several pulsars have masses around this value}. To better understand this result, we must investigate the behavior of the nuclear potential, which depends on the attractive scalar $\sigma$ meson and the repulsive vector $\omega$ meson.

The relativistic version of the Yukawa potential (Eq.~\ref{yuk2}) is~\cite{lopesPRD}:

\begin{equation}
 U(n) = g_\omega \omega_0 - g_s\sigma_0   
\end{equation}
where instead of the dependence on the distance ($r$), we let the dependence on the number density ($n$)~\footnote{Remembering that $n~\propto~k_F^3$ and in natural units $k_F~\propto~ 1/r$.}. From Eq.~\ref{sigma1} and Eq.~\ref{omega1} we can see that at low densities (and therefore, large distances),$M^* >> k_F$, and both fields grow proportional to $k_F^3$. The potential will be attractive or repulsive depending on which is larger, $G_S$ or $G_V$. However, at high densities, the $\omega$ field still grows proportional to $k_F^3$, but the $\sigma$ field grows proportional to $M^* k_F^2$~\footnote{Additionally, $M^*$ decreases with $k_F$, causing an extra slowing in the increase of the $\sigma$ field}. Therefore, the vector field dominates over the scalar one, even if $G_S~>G_V$.

To fully understand the behavior of the Yukawa potential, a numerical analysis of it with different values of $G_S$ and $G_V$ is needed. This is, nevertheless, left as an exercise for the reader.

\begin{center}
\begin{table}[h]
\begin{center}
\begin{tabular}{cccc}
\toprule
\textbf{Set}	& {$M_{max}/M_\odot$}	& {$R_{max}$~(km)} & {$R_{1.4}$~(km)} \\
Set A		& 0.71			& 8.87  & -\\
Set B		& 1.55			& 12.03  & 14.76\\
Set C		& 2.01			& 14.86  & 20.39\\
Set D		& 2.50			& 17.70  & 24.23\\
Set E		& 1.62			& 8.26   & 9.11\\
Set F		& 1.96			& 10.91  & 14.28\\
Set G		& 2.46			& 15.47  & 22.21\\
\toprule
\end{tabular}
\caption{Neutron stars' macroscopic properties for different values of $G_S$ and $G_V$.\label{T2}}
\end{center}
\end{table}
\end{center}

For Set B to Set D, where only the $\omega$ meson is present, the EOS becomes stiffer and stiffer as the density grows. Consequently, both the masses and radii grow. However, for Set E to Set G, the EOS presents a competition at low density and eventually becomes stiffer due to the vector meson dominance. This is especially true for Set E, which has $G_S~>~G_V$. At low densities, the EOS is even softer than free neutron matter. 
Eventually, it becomes stiffer, producing a maximum mass of around 1.6 $M_\odot$, but the soft EOS at low densities makes the radii much lower.

This gives us some hints about the relation between neutron stars and their EOSs.
A stiffer EOS will produce massive stars, but a softer EOS at low and moderate densities will produce more compact objects. \\

\section{First Nuclear Constraints}

As pointed out earlier, an effective nuclear model must ultimately describe, as accurately as possible,the atomic nuclei.

Based on decades of nuclear physics, the main features that a model of nuclear physics must satisfy can be summarized as~\cite{Glenbook,Lilley2001}:

\begin{itemize}
    \item Attractive and short ranged.

    The nuclear force is essentially attractive; otherwise, the existence of atomic nuclei would be impossible. Moreover, the nuclear force virtually goes to zero for $r~>2.5$ fm.

    \item Repulsive core.

    From the study of the scattering of $\alpha$ particles, the radius of the atomic nuclei can be experimentally inferred. It was found to be proportional to  $A^{1/3}$, where $A$ is the mass number of atomic nuclei:

    \begin{equation}
        r~\approx~ 1.16 A^{1/3}~ \mbox{fm}, \label{nucleiradius}
    \end{equation}
    The volume is therefore proportional to $A$, which implies that the density is constant. This led us to the concept of the saturation of the nuclear force, where the nuclear interaction must become repulsive at low distances.

    \item Charge independence. 
    
    The force between two protons is the same as the force between two neutrons. Moreover, the force between a proton and a neutron is "almost" the same if they form a state with the same spin. 

    \item Spin dependence.
    
    The nuclear force depends on the spin. For instance, the only known bound state of two nucleons is the deuteron. The deuteron has spin 1. $p+n$ with spin-0 does not form a bound state. In the same sense, $n+n$ and $p+p$ can only have spin 0 due to Pauli's exclusion principle. Effectively, this results in the average force between a neutron and a proton inside the nucleus being greater than the force between two identical nucleons. \\
\end{itemize} 

\subsection{Semi-Empirical Mass Formula}

The semi-empirical mass formula (SEMF), or Bethe–Weizsacker mass formula, gives us an approximation to the mass of an atomic nucleus from its number of protons ($Z$) and neutrons ($N$)~\cite{Bethe1936}:

\begin{equation}
    M = Z.M_p + N.M_n + B, \label{semf1}
\end{equation}
where $M_p$ is the mass of the proton, $M_N$ is the mass of the neutron, and $B$ is the total binding energy of the nucleus, which is negative, expressed as:

\begin{equation}
 B =  -a_V A +a_S A^{2/3} + a_C \frac{Z}{A^{1/3}} + a_A \frac{(N -Z)^2}{A} , \label{semf2}
\end{equation}
where $a_V$, $a_S$, $a_C$, and $a_A$~\footnote{Sometimes another parameter, called pairing parameter, is also present in this formula, as can be seen in ref~\cite{Lilley2001}. However, this parameter can be absorbed into $a_V$ as done in ref.~\cite{Glenbook}.} are constants that must be inferred to fit experimental results.

A very important quantity is the so-called binding energy per nucleon ($B/A$),  which is the average amount of energy required to separate a single nucleon from the nucleus of a given atom. It is obtained by dividing the binding energy $B$ by the mass number $A$. Alternatively, it can also be calculated via the total energy or energy density of a system~\cite{debora-universe}:

\begin{equation}
B/A =  \bigg (\frac{E}{A} - M\bigg )  =  \bigg (\frac{\epsilon_0}{n_0} - M\bigg ) . \label{BoA}
\end{equation}
where the subscript indicates that it must be calculated at the saturation density~\footnote{The saturation density, sometimes called saturation point ($n_0$), is the density where the binding energy per nucleon has a minimum. The reader must be able to show that, for thermodynamic consistency, the pressure must vanish at this point. However, this point does not represent the minimum of the Yukawa potential $U(n)$. The reader must reflect on this statement.  }.

The four terms in Eq.~\ref{semf2} are related, respectively, to the volume, surface, electric charge, and symmetry of the nucleus:

\begin{itemize}
    \item Volume term.

    This term comes directly from the saturation of the nuclear force. As the nuclear force saturates, the binding energy per nucleon becomes independent of $A$. Moreover, it is independent of $Z$ due to the charge independence.

    \item Surface correction.

    In a finite nucleus, nucleons closer to the surface will interact with fewer nucleons than nucleons close to the center of the nucleus.

    \item Coulomb term.

    Related to the electrostatic repulsion between the protons in the atomic nuclei.

    \item Symmetry term.

    Due to the Pauli exclusion principle,  nuclei with the same number of protons and neutrons are tightly bound compared to those with asymmetric proportions. However, the Pauli component alone is not enough to explain the strong tendency of equal numbers of protons and neutrons. The spin dependence also plays a role.

\end{itemize}

The standard values of $a_V$, $a_S$, $a_C$, and $a_A$ found in modern literature (see ref.~\cite{Glenbook,Lilley2001,Micaela2017,Dutra2014} and the references therein) are:

\begin{eqnarray}
  a_V = 15.7-16.5~\mbox{MeV}, \quad a_S =  17-19~\mbox{MeV}, \nonumber \\
  a_C =0.66-0.72~\mbox{MeV}, \quad a_A = 30-35~\mbox{MeV}. \label{valuesSEMF}
\end{eqnarray} \\

\subsection{The QHD and the Nuclear constraints}

We now can fix the coupling constants fo the QHD in order to reproduce the results coming from the SEMF. Let us begin with the concept of infinite symmetric nuclear matter. In this approximation, the nuclear matter is composite by a very large number of protons and neutrons in equal quantities. 
In this scenario, there is no boundary; therefore, no surface. The term related to the surface correction can be disregarded. Moreover, the Coulomb term is related to the electromagnetic force, and therefore, the QHD has no information about it. Finally, in symmetric nuclear matter, the symmetry term does not contribute. As $A~\to \infty$, from Eq.~\ref{semf2} we simply have:

\begin{equation}
 B/A = -a_V   
\end{equation}

Now we calculate $B/A$ from QHD using Eq.~\ref{BoA}, and determine the coupling constants to match the value of $a_V$ within the range presented in Eq.~\ref{valuesSEMF}.

The second constraint comes from Eq.~\ref{nucleiradius}. A radius of $1.16A^{1/3}$ fm implies a nuclear saturation density $n_0$ around 0.15 fm$^{-3}$. Modern values lie in the range\cite{Micaela2017,Dutra2014}.:

\begin{eqnarray}
n_0 =0.148-0.170~ \mbox{fm}^{-3}, \label{saturationdensity}    
\end{eqnarray}

Both constraints are related to symmetric nuclear matter at saturation density. Therefore, we need a Lagrangian that presents protons and neutrons. We postulate:

\begin{eqnarray}
\mathcal{L} = \sum_B\bar{\psi}_B[\gamma^\mu(i\partial_\mu -g_{B\omega}\omega_\mu) - (M_B - g_{Bs}\sigma)]\psi_B \nonumber \\+\frac{1}{2}(\partial^\mu\sigma\partial_\mu\sigma - m_s^2\sigma^2)    - \frac{1}{4}\Omega^{\mu\nu}\Omega_{\mu\nu} + \frac{1}{2}m_\omega^2\omega^\mu\omega_\mu . \label{sigma-omega-L2},
\end{eqnarray}
where the subscript $B =n,p$ runs over protons and neutrons. 
In this case, we have:

\noindent $M_p = M_n = M_N$, $g_{p\omega} = g_{n\omega} =g_{N\omega}$, and $g_{ps} = g_{ns} = g_{N\sigma}$, where the capital $N$ means nucleon.

The energy eigenvalue and chemical potential of each nucleon in MFA give:

\begin{eqnarray}
E_B = \mu_B =  \sqrt{M_B^{*2} + k_{FB}^2} + g_{B\omega}\omega_0
\end{eqnarray}

The total EOS also takes into account the individual amount of protons and neutrons:

  \begin{equation}
  \epsilon = \frac{8\pi}{(2\pi)^3}\sum_B \int_0^{k_{FB}} [\sqrt{M_B^{*2} +k^2}]  k^2 dk + \frac{1}{2}m_s^2\sigma^2 + \frac{1}{2}m_\omega^2\omega_0^2, \nonumber
 \end{equation}

 \begin{equation}
  p = \frac{1}{3\pi^2}\sum_B\int_0^{k_{FB}}\frac{k^4dk}{\sqrt{M_B^{*2} +k^2}}   - \frac{1}{2}m_s^2\sigma_0^2 + \frac{1}{2}m_\omega^2\omega_0^2 ,\label{psoNP}
 \end{equation}

The mesonic field, on the other hand, depends simultaneously on the amount of protons and neutrons:

\begin{equation}
\sigma_0 = \sum_B\frac{g_{Bs}}{(m_s)^2} n_{B}^S . \label{sigma2}, \nonumber
\end{equation}

\begin{equation}
\omega_0 = \sum_B\frac{g_{B\omega}}{(m_\omega)^2} n_{B} . \label{mesonicfields2},
\end{equation}

Finally, symmetric nuclear matter implicates $n_p = n_n \to k_{Fp} = k_{Fn}$.

In Tab.~\ref{T3} we display three sets able to satisfy these two constraints simultaneously. As can be seen, in the $\sigma-\omega$ model, the values of $G_S$ must lie in the range 14-16 fm$^2$. Consequently, the values of $G_V$ lies between 10-13 fm$^2$. As we increase $G_S$, we must increase $G_V$ to obtain $B/A$ in the range 15.8 - 16.3 MeV. However, a higher value of $G_V$ reduces the value of $n_0$. If we keep increasing $G_S$,  the value of $n_0$ will eventually lie outside the range 0.148 - 0.170 fm$^{-3}$.

\begin{figure*}[h!]
\begin{tabular}{ccc}
\centering 
\includegraphics[scale=.58, angle=270]{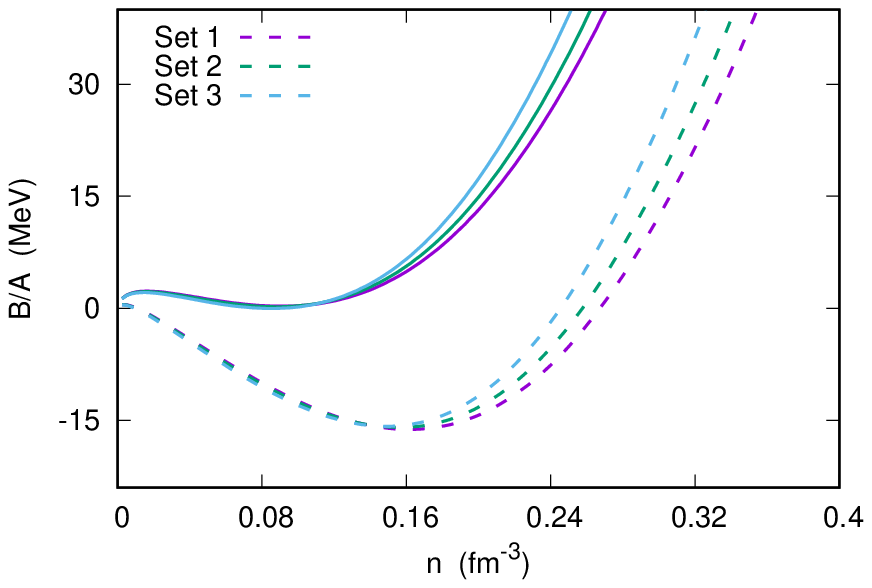} &
\includegraphics[scale=.58, angle=270]{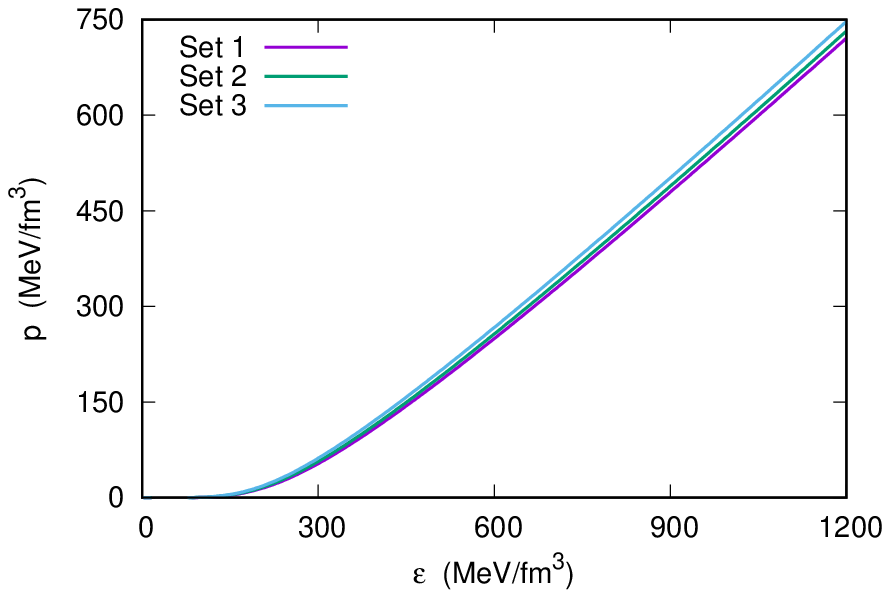} \\
\includegraphics[scale=.58, angle=270]{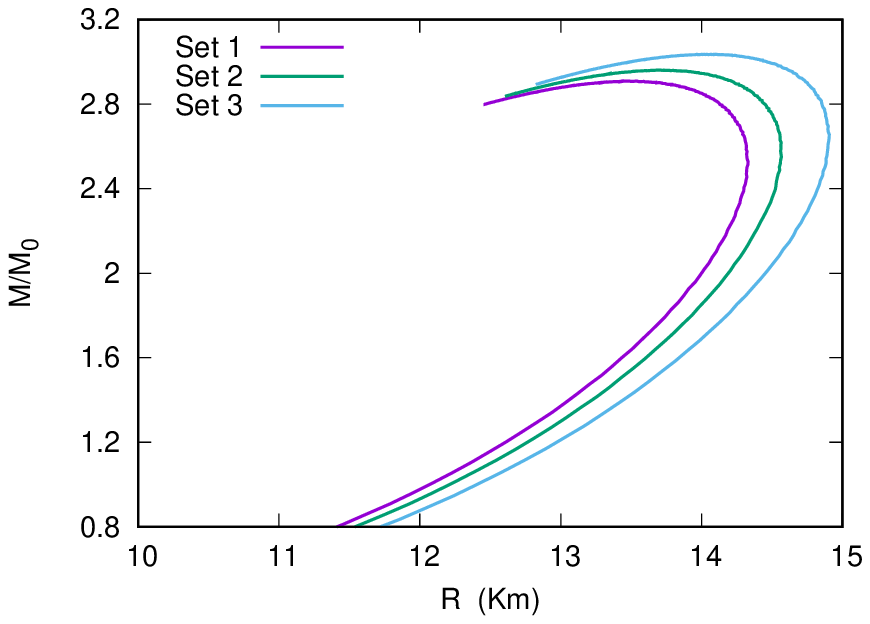} &
\end{tabular}
\caption{ (\textbf{a}) $B/A$,  (\textbf{b}) EOSs, and (\textbf{c}) OV solutions for minimally constrained models. Solid lines represent pure neutron matter, and dashed lines represent symmetric matter. \label{F4}}
\end{figure*}

In Fig.~\ref{F4}, we display the binding energy per nucleon for different values of $G_S$ and $G_V$  
for symmetric nuclear matter (dashed lines) and pure neutron matter (solid lines). 
 For pure neutron matter, we also display the corresponding EOSs and OV solutions. The higher the value of $G_V$, the higher our maximum mass. As can be seen, our simple model was able to generate neutron stars with maximum masses around $3.0~M_\odot$. \\

\begin{center}
\begin{table}[h]
\begin{center}
\begin{tabular}{ccccc}
\toprule
\textbf{Set}	& {$G_S$ (fm$^2$)}	& {$G_V$~(fm$^2$)} & {$B/A$ (MeV)} & $n_0$  (fm$^{-3}$)\\
Set 1		& 14.50			& 10.95  & -16.22 & 0.162 \\
Set 2		& 15.00			& 11.40  & -15.94 & 0.156\\
Set 3		& 15.75			& 12.06  & -15.81 & 0.149\\
\toprule
\end{tabular}
\caption{Different sets able to reproduce the constraints related to the binding energy per nucleon ($B/A$) and the saturation point $(n_0)$. \label{T3}}
\end{center}
\end{table}
\end{center}

\section{Toward a realistic neutron star description I: chemical equilibrium and neutron star's crust}

Despite satisfactorily fulfilling the constraints related to the binding energy per nucleon and the nuclear saturation point, our model, developed in the last section, still cannot be considered a minimally realistic model of neutron stars. \\

\subsection{Chemcial equilibrium}

The first failure of our model lies in the fact that we considered neutron stars as made of pure neutron matter. Neutrons, however, are unstable. Free neutrons decay into protons+electrons+neutrinos~\footnote {Technically, anti-neutrinos.} in about 14 minutes. In neutron stars' interior, neutrons are bound by gravity; therefore, we expect that neutron stars are made up of a majority of neutrons, but not pure neutron matter. Some must decay into protons, electrons, and (anti-)neutrinos.

\begin{equation}
   n = p +e +\bar{\nu}  .\label{pnev}
\end{equation}
Neutrinos are very weakly interacting and can be disregarded. Electrons are fermions of spin-1/2 and also obey the Dirac equation. Therefore, their energy eigenvalue and chemical potential are:

\begin{equation}
 E_e = \mu_e = \sqrt{m_e^2 +k_{Fe}^2}.  \label{ecp} 
\end{equation}

Chemical equilibrium, therefore, implies:

\begin{equation}
  \mu_n = \mu_p +\mu_e, \label{chemiel} \\
  \end{equation}
  \begin{eqnarray}
    \sqrt{M_n^{*2} +k_{Fn}^2} +g_{n\omega}\omega_0 =  \sqrt{M_p^{*2} +k_{Fp}^2} +g_{p\omega}\omega_0 +  \sqrt{m_e^2 +k_{Fe}^2}. \label{chemequi}
 \end{eqnarray}

 As neutron stars are electrically neutral, we must impose $n_p =n_e \to k_{Fp} =k_{Fe}$. Moreover, as $g_{n\omega} =g_{p\omega}$, the $\omega$ mesons does not contribute to the chemical equilibrium. Finally, as $m_e$ = 0.51 MeV, implying in $m_e << M^{*}_n$, and $M_p^* =M_n^*$, the Fermi momentum of electrons and protons can be given in a good approximation by:

 \begin{equation}
k_{Fp} =k_{Fe}~\approx~ \frac{k_{Fn}^2}{2\sqrt{M_n^{*2} +k_{Fn}^2}} . \label{electronsed}
 \end{equation}

 In the high density limit, $k_{Fp} =k_{Fn}/2 \to n_p = n_n/8$, independently of the values of $G_S$ and $G_V$~\footnote{As we will see, when the $\rho$ field is present, the proton fraction can reach significantly large values. }.

 Finally, the energy density and pressure of the electrons are:

\begin{equation}
 \epsilon_e = \frac{1}{\pi^2}\int_0^{k_{Fe}}\sqrt{m_e^{2} +k^2} k^2 dk ,   \label{edsigma2}
\end{equation}

\begin{equation}
  p_e = \frac{1}{3\pi^2}\int_0^{k_{Fe}} \frac{k^4 dk}{\sqrt{m_e^{2} +k^2}}.  \label{electronsp}
\end{equation} \\

 \subsection{The crust}

The second issue is the neutron star's crust. The QHD as used here is only valid if the density is high enough to neglect the surface and Coulomb terms. 
As the density decreases towards the surface, these effects cannot be ignored anymore. We are in a region called the neutron star crust. 
The neutron star crust can be divided into two parts~\cite{Glenbook,debora-universe,lopes2021EPL}:\\

\begin{itemize}
    \item Outer crust: Region with density $n~<10^{-4}$ fm$^{-3}$. In this region, the effects of nuclear physics are almost irrelevant, and the nuclei form a perfect crystal with a single nuclear species. In the low-density limit, the ground state is a crystal lattice of $^{56}$Fe with negligible —but increasing— pressure. For higher densities, the matter is a plasma of
nuclei and electrons which form a nearly uniform Fermi gas, with the degeneracy pressure of electrons and a small lattice pressure. For densities above $10^{-7}$ fm$^{-3}$, $^{62}$ Ni replaces $^{56}$ Fe as the ground state, and then is replaced by the extreme neutron-rich nuclei $~^{78}$ Ni.  In even higher densities, a crystallized phase can occur, with the presence of even heavier nuclei, as the $^{86}$Kr and $^{124}$Mo.~\cite{Fatina2020}.\\

\item Inner Crust: The inner crust lies in the range $10^{-4}~\lesssim~n~\lesssim10^{-1}$ fm$^{-3}$. In this region, neutrons begin to ``drip” out of the nuclei. The ground state consists of a lattice of nuclei immersed in a pure neutron gas, in addition to the electron gas. As the density increases, the nuclei dissolve into a matter consisting of a uniform liquid of neutrons with a small fraction of protons and electrons.  It is also possible that in the densest layers of the crust, the Coulomb energy becomes comparable in magnitude to the net nuclear binding energy. The matter thus becomes frustrated and can arrange itself into various exotic configurations as observed in complex fluids. This is called the pasta phase. Another possibility is the presence of neutron superfluidity. \\

\end{itemize}

The study of the neutron star crust is beyond the scope of the present work. We use here the BPS EOS for the outer crust~\cite{BPS}, and the BBP EOS for the inner crust~\cite{BBP}. Additional discussion can be found in these original papers, as well as in Ref.~\cite {Glenbook} and the references therein. Another important work is ref.\cite{Fortin2016}, where the effects of different approaches to the neutron stars' crust are studied in detail. The possible existence of an exotic phase, as the pasta phase and superfluidity, is discussed in ref.~\cite{Chamel2013}.

\begin{figure*}[h!]
\begin{tabular}{ccc}
\centering 
\includegraphics[scale=.58, angle=270]{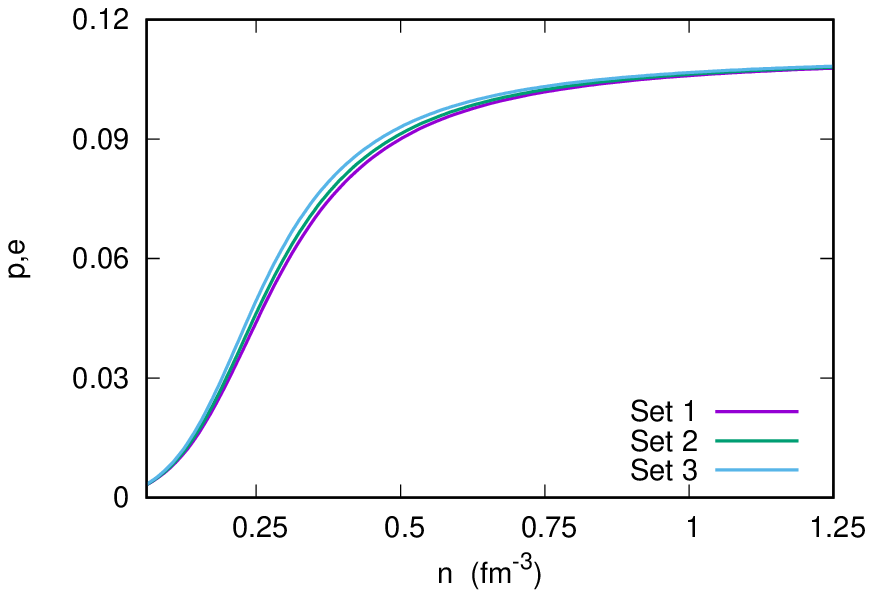} &
\includegraphics[scale=.58, angle=270]{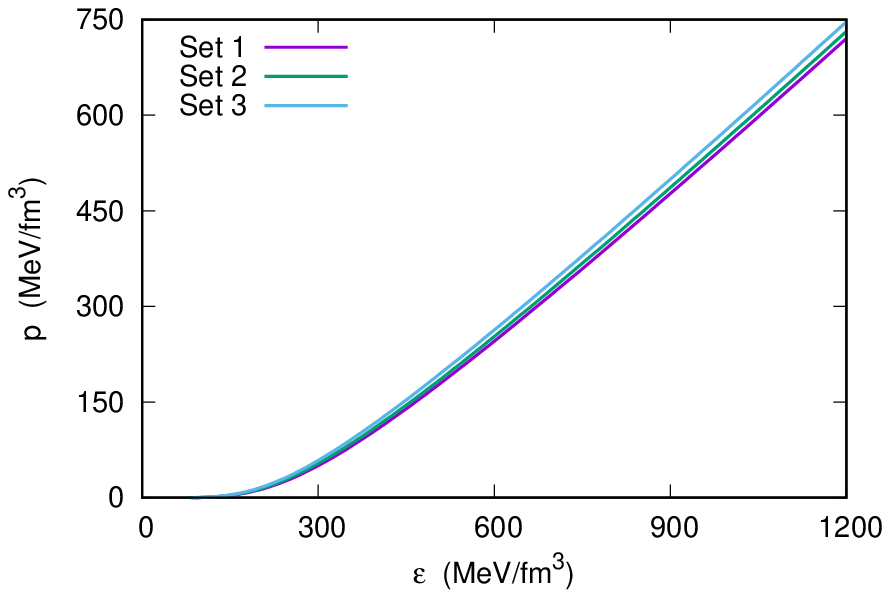} \\
\includegraphics[scale=.58, angle=270]{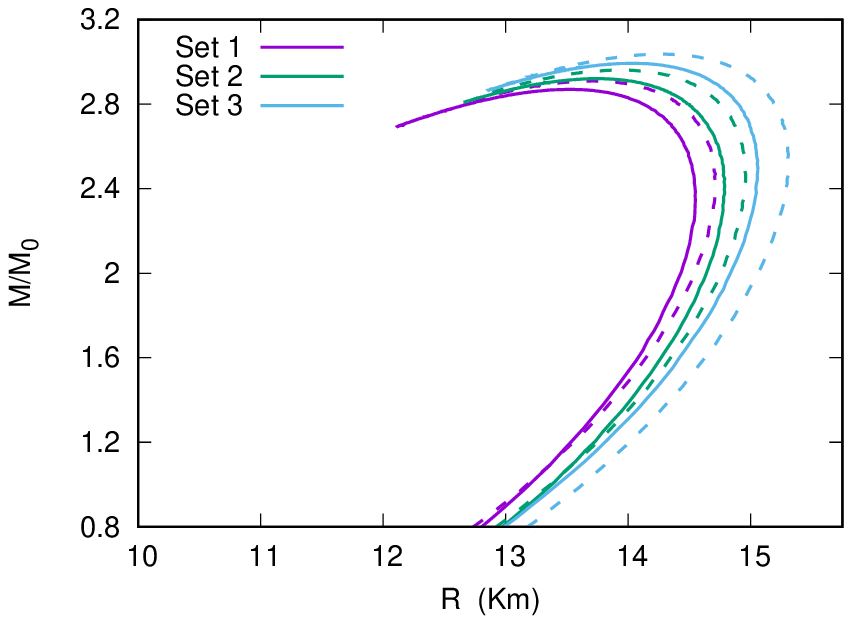} &
\end{tabular}
\caption{ (\textbf{a}) Proton and electron fractions. (\textbf{b}) EOSs for beta-stable matter. (\textbf{c}) OV solutions with the crust for beta-stable matter (solid lines) and pure neutron matter (dashed lines). \label{F5}}
\end{figure*}

In Fig.~\ref{F5}, we show the properties of beta-stable matter for Set 1 to Set 3. In {\bf (a)}, we present the proton and electron fraction $Y_p = n_p/n$, as a function of density. As can be seen, the relative number of protons and electrons grows with the density as expected, following Eq.~\ref{electronsed}. Furthermore, the proton fraction reaches an asymptotic limit of 1/8 of the neutron fraction, implying a fraction of 1/9 of the total number density. The EOSs presented in {\bf (b)} are visually identical to the EOSs for pure neutron matter presented in Fig.~\ref{F4}. The main differences are in the mass-radius relation obtained by solving the OV equations, which we display in {\bf (c)} for both: beta-stable matter (solid lines) and pure neutron matter (dashed lines), but now including the crust.

We can notice that the main effect of the crust is to increase the radii of the neutron stars. Moreover, when we compare pure neutron matter with beta-stable matter, we notice that beta-stable stars are smaller and have a lower maximum mass. These effects are due to the softening of the EOS caused by the onset of protons and electrons, which can occupy lower energy states due to the Pauli exclusion principle.

Crusted beta-stable matter within the $\sigma-\omega$ model was the state-of-the-art theory of neutron stars until the late 1970s~\cite{WALECKA1974,CANUTO1979NPA}. From this point on, all neutron stars analysed will have a crust.\\

\section{The vector-isovector $\rho$ meson and the $\sigma\omega\rho$ model }

The careful reader may have noticed that a term in the SEMF has not been discussed yet. The symmetry term, related to the constant $a_A$.

As pointed out earlier, the binding energy per nucleon for infinite symmetric nuclear matter is $B/A = -a_V$. On the other hand, the binding energy per nucleon for pure neutron matter, where $A = N$, according to Eq.~\ref{semf2} is:
$B/A = -a_V + a_A$. The parameter $a_A$ can therefore be calculated as the difference of the binding energy per nucleon of the pure neutron matter (PNM) in relation to the binding energy per nucleon of symmetric nuclear matter (SNM) at the saturation point:

\begin{equation}
  B/A \bigg |_{PNM} - B/A \bigg |_{SNM} = a_A.  \label{se1}
\end{equation}

The term $a_A$ is sometimes called the symmetry energy of the system. 
However, this nomenclature is not unique. In modern texts, the symmetry energy is defined as the second term in the expansion of $B/A$ in terms of the asymmetric parameter $\alpha$~\cite{Glenbook,LI2008PRep}.

\begin{equation}
   B/A =  -E_(n) + S(n)\alpha^2 + O(\alpha^4), \quad \mbox{with} \quad \alpha = \frac{n_n - n_p}{n_n +n_p},
\end{equation}
where

\begin{equation}
  S(n) = \frac{1}{2} \bigg (\frac{\partial^2 (\epsilon/n)}{\partial \alpha^2 } \bigg )  \bigg |_{\alpha =0} . \label{Sn}
\end{equation}

At the saturation density, we have $E(n_0)= a_V$, while $S(n_0) =S_0$ represents the symmetry energy of the system. To calculate this expression, we begin by defining the Fermi momentum of protons and neutrons in the symmetric matter $k_F = k_{Fp} = k_{Fn}$ in terms of the saturation density:

\begin{equation}
n  = n_p +n_n= 2\frac{k_F^3}{3\pi^2} \to k_F = \bigg ( \frac{3\pi^2n}{2} \bigg )^{1/3},
\end{equation}
where the factor 2 appears, as we are summing equal amounts of protons and neutrons.
Now, to take the asymmetry into account,  $k_{Fn}$ and $k_{Fp}$ can be written as a function of $\alpha$:

\begin{equation}
k_{Fp} = k_F(1 +\alpha)^{1/3}, \quad k_{Fn} = k_F(1 -\alpha)^{1/3}.
\end{equation}

We obtain the following:

\begin{equation}
  S(n) = \frac{k_F^2}{6\sqrt{M^{*2}_N + k_F^2}}.  \label{Sn2}
\end{equation}

Within our models, following Eq.~\ref{se1}, we obtain the values of the symmetry parameter $a_V$ = 21.5, 20.8, and  20.6 MeV for Set 1, Set 2, and Set 3, respectively. In the same sense, within Eq.~\ref{Sn2}, at the saturation density, we have: $S_0$ = 20.4, 19.7, and 19.5 MeV. As can be seen, independently of our approach to the symmetry energy, our values lie below the expected values constrained in the modern literature, $S_0 = 30 -35$ MeV~\cite{Dutra2014,Micaela2017}. 

 Another issue that must be noted is the presence of a minimum in the energy density per baryon for pure neutron matter presented in Fig.~\ref{F4} (a). This implies that pure neutron matter can form a bound state, a result that is experimentally ruled out.

We, therefore, need an interaction that can raise the symmetry energy of the system without changing the already constrained values of $B/A$ and $n_0$,  as well as prevent the existence of bounded pure neutron matter.
These tasks are accomplished by introducing an interaction able to differentiate protons from neutrons, the $\rho$ mesons triplet.

The $\rho$ mesons are vector-isovector mesons, ie., they are mesons of spin-1, and are also vector in the isospin space. On the other hand, the $\sigma$ and $\omega$ mesons are isoscalar, ie., they are scalar in the isospin space~\footnote{Alternatively, $\omega$ and $\sigma$ mesons belong to the U(1) group, while the $\rho$ mesons belong to the SU(2) group. See a discussion in ref.~\cite{griffiths_part}.}. The $\rho$ mesons, as the $\omega$, are also real mesons with masses of approximately 770 MeV~\cite{PDG2020}. The  Lagrangian in the $\sigma\omega\rho$ model reads~\cite{Glenbook}:

\begin{eqnarray}
\mathcal{L} = \sum_B\bar{\psi}_B[\gamma^\mu(i\partial_\mu -g_{B\omega}\omega_\mu - \frac{1}{2}g_{B\rho}\vec{\tau}\cdot \vec{\rho}_\mu) - (M_B - g_{Bs}\sigma)]\psi_B \nonumber \\+\frac{1}{2}(\partial^\mu\sigma\partial_\mu\sigma - m_s^2\sigma^2)    - \frac{1}{4}\Omega^{\mu\nu}\Omega_{\mu\nu} + \frac{1}{2}m_\omega^2\omega^\mu\omega_\mu \nonumber \\
- \frac{1}{4}{\bm{P^{\mu\nu}\cdot P_{\mu\nu}}} + \frac{1}{2}m_\rho^2(\vec{\rho}^\mu \cdot\vec{\rho}_\mu )\label{sigma-omega-rho},
\end{eqnarray}
where {$ \bm P_{\mu\nu}~\equiv$~} $\partial_\mu \vec{\rho}_\nu - \partial_\nu\vec{\rho}_\nu -g_\rho(\vec{\rho}_\mu \times \vec{\rho}_\nu)$, and $\vec{\tau}$ are the Pauli matrices~\footnote{Here, we use $\vec{\tau}$ instead of $\vec{\sigma}$ because it lies on the isospin space instead of the spin space. This notation is the same as presented in ref.\cite{Glenbook}.}.

In MFA, only the uncharged $\rho$ meson survives, due to the charge and baryon number conservation~\footnote{A more detailed discussion about the $\rho$ mesons in QHD can be seen in ref.~\cite{Glenbook} and references therein.}. Moreover, assuming that the baryon matter are static and in the ground state, only the temporal component of the $\rho_0^\mu$ survives, in analogy with the $\omega^\mu$ meson in MFA. We then have: 

\begin{equation}
   \vec{\rho}_\mu \to \rho_0~\equiv~\delta_0^\mu\delta_j^3\langle\rho_\mu^j \rangle. 
\end{equation}

Moreover, $g_{n\rho} = g_{p\rho} = g_{N\rho}$. The Lagrangian in MFA then reads:

\begin{eqnarray}
\mathcal{L} = \sum_B\bar{\psi}_B[\gamma^0(i\partial_0 -g_{N\omega}\omega_0 - \frac{1}{2}g_{N\rho}{\tau_3} {\rho}_0) -i\gamma^j \partial_j - M_B^*]\psi_B \nonumber \\-\frac{1}{2}m_s^2\sigma_0^2    + \frac{1}{2}m_\omega^2\omega_0^2
+ \frac{1}{2}m_\rho^2{\rho}_0^2\label{sigma-omega-rhoMF},
\end{eqnarray}

Now, applying E-L in Eq.~\ref{sigma-omega-rhoMF} relative to the nucleon Dirac field in MFA, we obtain for each nucleon:

\begin{equation}
  [\gamma^0(i\partial_0 - g_{N\omega}\omega_0 - \frac{1}{2}g_{N\rho}{\tau_3} {\rho}_0) - i\gamma^j\partial_j  - M^{*}_N]\psi = 0]. \label{Dirac3}  
\end{equation}

Using the quantization rules from Eq.~\ref{quantrules}:

\begin{eqnarray}
    \bigg [\left( \begin{array}{ll}
         1 & 0\\
        0 & -1 \end{array} \right ) \cdot (E_B- g_{N\omega} \omega_0 - \frac{1}{2}g_{N\rho}{\tau_3} {\rho}_0)~-~
         \left( \begin{array}{ll}
         0 & \vec{\sigma}\\
        -\vec{\sigma} & 0 \end{array} \right )\cdot \vec{k}~ \nonumber \\ -
        \left( \begin{array}{ll}
         1 & 0\\
        0 & 1 \end{array} \right )\cdot M^{*}_N \bigg]   \left( \begin{array}{l}
         u_A \\
         u_B \end{array} \right ) =0 .
        \label{DSOR}
\end{eqnarray}

Eq.~\ref{DSOR} is analogous to Eq.~\ref{DS1} with a effective mass $M^*$ and an effective energy $E_B^* = E_B - g_{N\omega} \omega_0 - \frac{1}{2}g_{N\rho}{\tau_3} {\rho}_0$. Therefore its solution is $E_B^* = \sqrt{M_N^{*2} +k_B^2}$, which produces the energy eigenvalue of:

\begin{equation}
E_B = \mu_B = \sqrt{M^{*2}_N + k_B^2} + g_{N\omega}\omega_0 +\frac{1}{2}g_{N\rho}{\tau_3} {\rho}_0    .\label{ESOR}
\end{equation}

Now, $\tau_3$ assumes +1 for protons and -1 for neutrons. Therefore, they energy eigenvalues are:

\begin{eqnarray}
  E_B = \mu_B  = \left\{ \begin{array}{ll}
         \sqrt{M^{*2}_N + k_B^2} + g_{N\omega}\omega_0 +\frac{1}{2}g_{N\rho} {\rho}_0  & \mbox{for protons};\\
      \sqrt{M^{*2}_N + k_B^2} + g_{N\omega}\omega_0 -\frac{1}{2}g_{N\rho} {\rho}_0  & \mbox{for neutrons}.\end{array} \right.  \label{energyrho}
\end{eqnarray}

With the Fermi-Dirac distribution, we can obtain the energy density of the protons, neutrons, and electrons. At the same time, the mesonic contribution of the Hamiltonian in MFA is given by $\langle \mathcal{H}\rangle  = - \langle \mathcal{L}\rangle$. We therefore have:

 \begin{eqnarray}
  \epsilon = \frac{1}{\pi^2} \sum_B\int_0^{k_{FB}} [\sqrt{M_N^{*2} +k^2} + g_{N\omega}\omega_0 +\frac{1}{2}g_{N\rho}\tau_3\rho_0]  k^2 dk \nonumber \\  + \frac{1}{2}m_s^2\sigma_0^2 - \frac{1}{2}m_\omega^2\omega_0^2 - \frac{1}{2}m_\rho^2\rho_0^2 + \frac{1}{\pi^2}\int_0^{k_{Fe}}\sqrt{m_e^{2} +k^2} k^2 dk. \label{endsor}
 \end{eqnarray}

Finally, the expected value of the $\rho$ field is obtained by imposing the energy density to be stationary at fixed baryon density, $\partial \epsilon/\partial \rho_0  = 0$, as done for the $\sigma$ and $\omega$ mesons: 

\begin{equation}
\bigg ( \frac{\partial \epsilon}{\partial \rho_0} \bigg ) = \frac{1}{\pi^2}\sum_B\int_0^{k_{FB}} \frac{\tau_3}{2}{g_N\rho}k^2 dk -m_\rho^2\rho_0  = \sum_B\frac{1}{2}g_{N\rho}\tau_3n_B  - m_\rho^2\rho_0 =0,  \label{estarho}
\end{equation}
which gives us:

\begin{equation}
\rho_0 = \sum_B\frac{g_{N\rho}}{(m_\rho)^2}\frac{\tau_3}{2} n_{B} = \frac{g_{N\rho}}{(m_\rho)^2}\frac{n_p -n_n}{2}\label{mesonicfields3}.
\end{equation}

From Eq.~\ref{mesonicfields3} it is clear that the field $\rho$ is negative in beta-stable matter, due to the much larger amount of neutrons compared to protons. Therefore, according to~\ref{energyrho}, the $\rho$ field increases the chemical potential of neutrons while reducing it for protons. Furthermore, for symmetric nuclear matter, $n_p =n_n$ and the $\rho$ field vanishes, ensuring that the constraints related to $B/A$ and $n_0$ are not affected.

Combining Eq.~\ref{estarho} with Eq.~\ref{endsor}, the energy density can be rewritten as:

 \begin{eqnarray}
  \epsilon = \frac{1}{\pi^2} \sum_B\int_0^{k_{FB}} [\sqrt{M_N^{*2} +k^2}]  k^2 dk \nonumber  + \frac{1}{2}m_s^2\sigma_0^2 + \frac{1}{2}m_\omega^2\omega_0^2 \\ + \frac{1}{2}m_\rho^2\rho_0^2 + \frac{1}{\pi^2}\int_0^{k_{Fe}}\sqrt{m_e^{2} +k^2} k^2 dk. \label{endsor2}
 \end{eqnarray}

 The pressure of baryons and electrons is given by Eq.~\ref{thermopressure}, while the pressure of mesons is $p_m$ = $\langle \mathcal{L}\rangle$. The total pressure is:

 \begin{eqnarray}
  p = \frac{1}{3\pi^2}\sum_B\int_0^{k_{FB}}\frac{k^4dk}{\sqrt{M_B^{*2} +k^2}}   - \frac{1}{2}m_s^2\sigma_0^2 \nonumber \\ + \frac{1}{2}m_\omega^2\omega_0^2  
+ \frac{1}{2}m_\rho^2\rho_0^2 +  \frac{1}{3\pi^2}\int_0^{k_{Fe}}\frac{k^4dk}{\sqrt{m_e^{2} +k^2}} .\label{p-sor}
 \end{eqnarray}

Now that the $\rho$ field is properly introduced, we study how it affects the symmetry energy. The value of $a_A$ is still calculated as the difference between the binding energy per nucleon of pure neutron matter in comparison with symmetric nuclear matter; however, the analytical expression for $S(n)$ derived from Eq.~\ref{Sn} yields:

\begin{equation}
  S(n) = \frac{n}{8} \bigg (\frac{g_{N\rho}}{m_\rho} \bigg )^2  + \frac{k_F^2}{6\sqrt{M^{*2} + k_F^2}}. \label{Snrho}
\end{equation}

 As can be seen, the $\rho$ field has a strong impact on the symmetry energy.
 
 Regarding beta-stable matter, it is clear from Eq.~\ref{energyrho} that the $\rho$ field affects the proton fraction. Because the $\rho$ field increases the neutron chemical potential, it becomes energetically favorable to convert more neutrons into protons+electrons. The higher the value of $g_{N\rho}$, the higher the amount of protons. \\

 \subsection{Numerical Results}
 
We begin here by defining $G_\rho$ as analogous to $G_V$ and $G_S$: 

\begin{equation}
 G_\rho = \bigg (\frac{g_{N\rho}}{m_\rho} \bigg )^2 .
\end{equation}

Qualitatively, the effects of $G_\rho$ are independent of $G_S$ and $G_V$, therefore, we use here only Set 2 from Tab.~\ref{T3} for these parameters. The values of $G_\rho$ and their physical quantities are presented in Tab.~\ref{T4}.
From our discussion about nuclear constraints, only Set2c can be faced as realistic.

\begin{center}
\begin{table}[h]
\begin{center}
\begin{tabular}{cccc}
\toprule
\textbf{Set}	& {$G_\rho$ (fm$^2$)}	& {$S_0$~(MeV)} & {$a_A$ (MeV)} \\
Set 2a		& 0.00			& 19.7  &  20.8 \\
Set 2b		& 1.87			& 27.0  &  28.1 \\
Set 2c		& 3.38			& 32.8  &  33.9 \\
Set 2d		& 5.00          & 39.1  &  40.1 \\
\toprule
\end{tabular}
\caption{Symmetry energy parameters for different values of $G_\rho$ within Set 2. \label{T4}}
\end{center}
\end{table}
\end{center}

\begin{figure*}[h!]
\begin{tabular}{ccc}
\centering 
\includegraphics[scale=.58, angle=270]{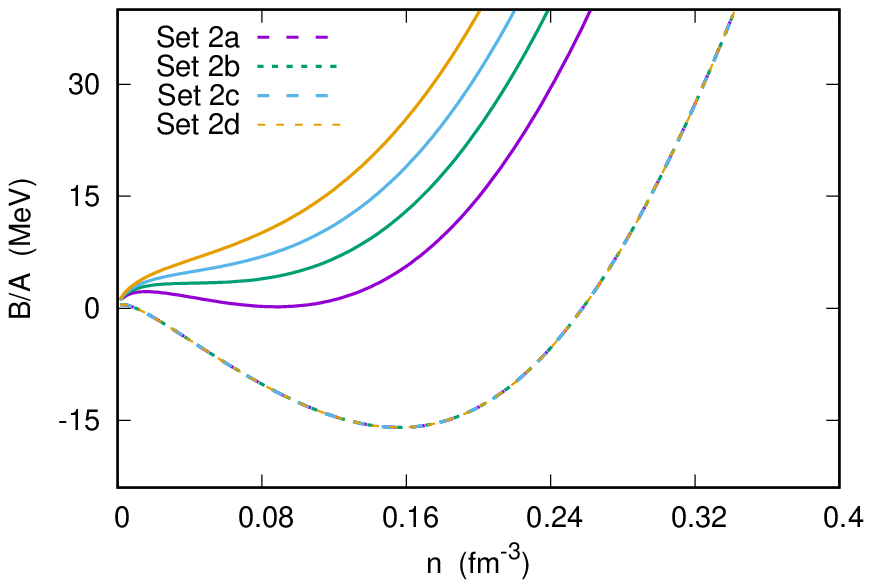} &
\includegraphics[scale=.58, angle=270]{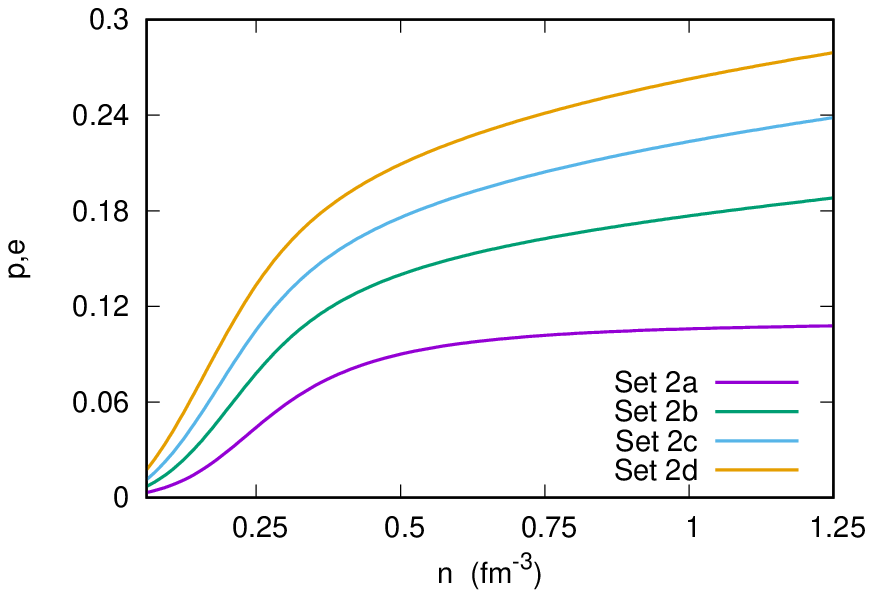} \\
\includegraphics[scale=.58, angle=270]{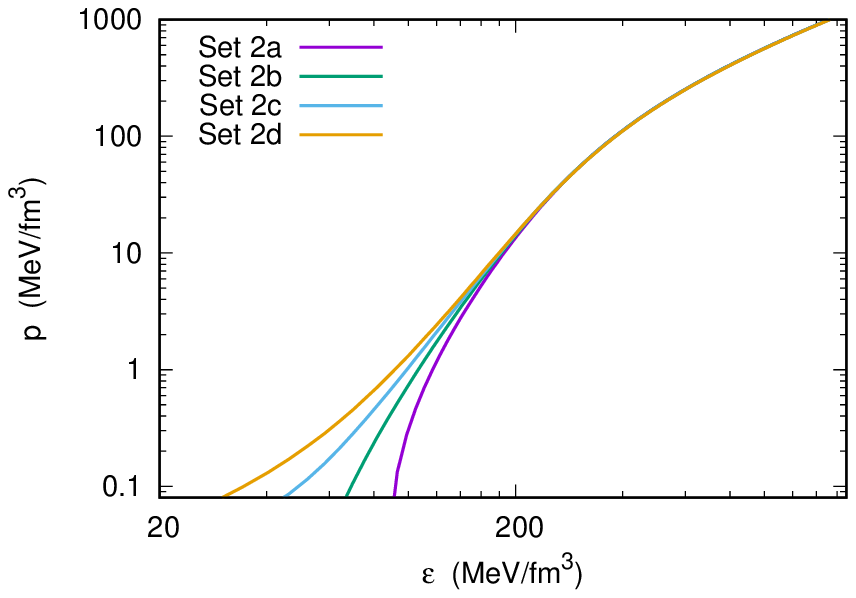} &
\includegraphics[scale=.58, angle=270]{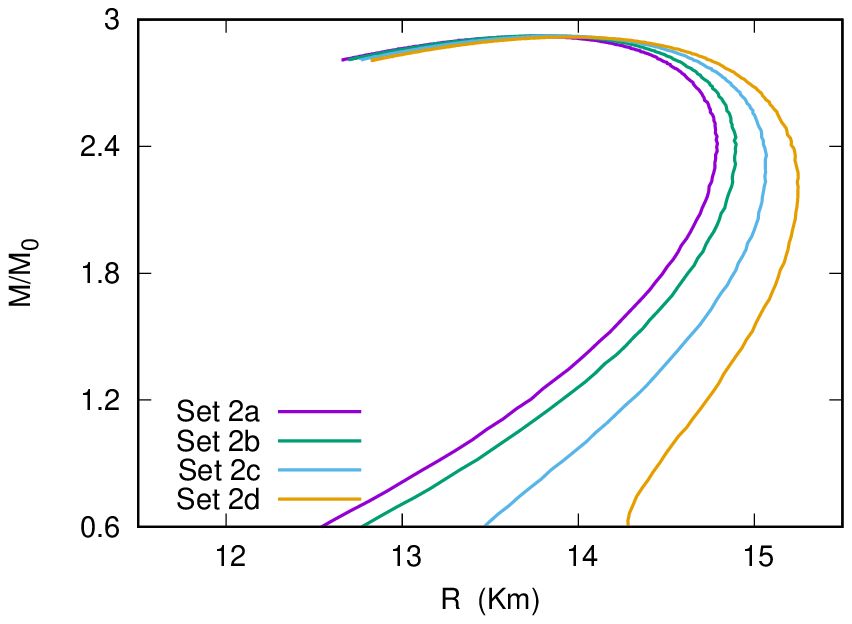} \\
\end{tabular}
\caption{Influence of the $\rho$ field in nuclear matter: (\textbf{a}) Binding energy per nucleon for symmetric matter (dashed lines) and pure neutron matter (solid lines) (\textbf{b}) Proton and electron fractions for different values of $G_\rho$ in beta-stable matter. (\textbf{c}) EOSs. (\textbf{d}) OV solutions.\label{F6}}
\end{figure*}

In Fig.~\ref{F6}, we display the properties within the $\sigma\omega\rho$ model.
In {\bf (a)}, we show the binding energy per nucleon for symmetric matter (dashed lines) and pure neutron matter (solid lines). As can be seen, the results for symmetric matter are all degenerate once the $\rho$ field vanishes. For pure neutron matter, as we increase the value of $G_\rho$, we also increase $B/A$, increasing $a_A$.  The minimum in $B/A$ for pure neutron matter is also removed when $\rho$ field is strong enough; in our case, Sets 2c and 2d.

In {\bf (b)}, we show the proton and electron fractions in beta-stable matter. Increasing $G_\rho$, increases the fraction of protons. In {\bf (c)}, we show the EOSs for beta-stable matter. Visually, the EOSs are very similar. In order to help us visualize their differences, they are presented on a logarithmic scale. We can see that at low densities, larger values of $G_\rho$ produces stiffer EOSs. However, at large densities the EOSs become degenerate due to the $\omega $ dominance~\footnote{The $\omega$ meson grows proportional to $n_p+n_n$, while the $\rho$ meson grows with $1/2(n_p -n_n)$. Besides, we have $G_V >2 G_\rho$.}.

As pointed out earlier, as we have a stiffer EOS at low densities, but a degenerate state at higher densities, we expected that the maximum mass of neutron stars would be very similar, while the radii of lower masses should grow with $G_\rho$.
This is exactly what happens, and it is displayed in {\bf (d)}. \\

\section{Toward a realistic neutron star description II: muons and non-linear $\sigma$ coupling.}

\subsection{Muons} 

Let us consider the following decay:

\begin{equation}
    n\to p +\mu +\bar{\nu}_\mu. \label{muonce1}
\end{equation}

Of course, such decay cannot happen at the neutron's rest frame, once $m_\mu$ = 105.66 MeV. 
 However, as the density increases, the chemical potential increases as well, and such decay becomes available. Imposing chemical equilibrium, and disregarding the neutrino, we have:

\begin{equation}
  \mu_n = \mu_p +\mu_\mu.  
\end{equation}

Now, comparing with Eq.~\ref{chemiel}, we have:

\begin{eqnarray}
   \mu_\mu =\mu_e \label{muon=electron}
\end{eqnarray}

Muons are also fermions with spin-1/2 that obey the Dirac equation: Their chemical potentials are identical to those of electrons, but their masses (Eq.~\ref{ecp}).
Due to the Pauli exclusion principle, it can become energetically favorable to produce muons instead of electrons if the electron chemical potential becomes too large. From Eq.~\ref{muon=electron}, we can obtain an equation for the Fermi momentum of the muon:

\begin{eqnarray}
  k_{F\mu}^2 = \sqrt{m_e^2 +k_{Fe}^2} -m_\mu^2  .
\end{eqnarray}

Therefore, from the definition of fermion number density (Eq.~\ref{numberd}):

\begin{eqnarray}
  n_\mu  = \left\{ \begin{array}{ll}
         0 & \mbox{if $k_{F\mu}^2~\leq~0$ };\\
      {k_{F\mu}^3}/{(3\pi^2)}  & \mbox{if $k_{F\mu}^2~>~0$ }.\end{array} \right.  \label{muond}
\end{eqnarray}

Calculations with various models suggest that muons appear at densities slightly below the saturation point  (around 0.11 to 0.12 fm$^{-3}$, depending on the model). Indeed, the standard model of neutron stars consists of protons, neutrons, electrons, and muons~\footnote{Abbreviated as $npe\mu$ matter.} in chemical equilibrium and with zero electric charge net.

\begin{eqnarray}
    \mu_n = \mu_p +\mu_e , \nonumber \\
    \mu_\mu =\mu_e, \nonumber \\
    n_p = n_e +n_\mu. \label{npeu}
    \end{eqnarray} \\

\subsection{Non-linear $\sigma$ model and the incompressibility of the nuclear matter}

The $\sigma\omega\rho$ model is able to correctly predict the constraints coming from the SEMF, at least for the infinite nuclear matter. However, the SEMF was proposed in the 1930s~\cite{Bethe1936}. Naturally, new constraints appear as our knowledge about nuclear physics grows. 

One of these new constraints is the incompressibility of nuclear matter, $K_0$.
This constraint came from the discovery of the so-called giant monopole resonance (GMR) in the late 1970s~\cite{Harakeh1977}. GMR is a collective excitation, in which the nuclei perform a pure radial oscillation (hence the name monopole), also called the 'breathing mode' of the nucleus~\cite{BLAIZOT1980}.

For any thermodynamic system, the compressibility modulus, $\chi$, is given by~\cite{Blundell_CTP}:

\begin{equation}
    \chi = -\frac{1}{V}\bigg (\frac{\partial V}{\partial p} \bigg ).
\end{equation}

    Now, as we done before, we define: $E = \epsilon V$ and $V = N/n$. With the help of Eq.~\ref{thermopressure}, we obtain~\cite{BLAIZOT1980,Schmitt2010}:

    \begin{equation}
     \chi = \frac{1}{n} \bigg ( \frac{\partial p}{\partial n} \bigg )^{-1} \label{chi}
    \end{equation}

    Now, the incompressibility of symmetric nuclear matter is defined as:

    \begin{equation}
        K_0 = 9n^2 \bigg (\frac{\partial^2 (\epsilon/n)}{\partial n^2} \bigg ) \bigg |_{n =n_0} \label{Kexp}
    \end{equation}

    But, by the definition of pressure (Eq.~\ref{thermopressure}):

    \begin{equation}
       K_0 = 9n^2 \bigg [\frac{\partial }{\partial n} \bigg (\frac{p}{n^2} \bigg ) \bigg ] =9  \bigg (\frac{\partial p}{\partial n} - \frac{2p}{n}\bigg)  \bigg |_{n =n_0}\label{k0} .
    \end{equation}

Now, at the saturation density, the pressure vanishes. Comparing Eq.~\ref{chi} with Eq.$\ref{k0}$, we can write the nuclear incompressibility as a function of the classical thermodynamical compressibility at the saturation density:

\begin{equation}
   K_0 = \frac{9}{n_0\chi}. 
\end{equation}
    It is clear from the discussion above that the nuclear incompressibility, unlike other quantities as the binding energy per nucleon or the symmetry energy that can be calculated at any desired density, is only meaningful at the saturation point.
    
    At saturation density, incompressibility can be viewed as an " effective spring constant" of the nucleus. The higher its value, the harder the "spring", implying a stiffer EOS, at least, for densities close to the saturation point~\footnote{Indeed, the incompressibility cannot be considered as the parameter that defines the EOS stiffness. For example, the NL3 parameterization of QHD~\cite{Lala1997} is stiffer than the GM1 parametrization~\cite{GlenPRL}, even though the GM1 has a higher value of $K_0$.}.

    The analyses of the giant monopole resonances pointed out that the value of incompressibility lies around 250 MeV~\cite{BLAIZOT1980,Harakeh1977}. Modern constraints~\cite{Dutra2014,Micaela2017} bounds $K_0$ to:

    \begin{equation}
        K_0 = 240~\pm~20~\mbox{MeV}. \label{k0values}
    \end{equation}

The big problem is that models within the $\sigma\omega\rho$ model, as our Set 2c predicts $K_0~\approx~550$ MeV, which is more than twice the commonly accepted value.

There is another constraint that our $\sigma\omega\rho$ model does not satisfy: the effective nucleon mass at saturation density, sometimes called the Dirac mass, $M_N^*/M_N$. Its value can be derived from a non-relativistic analysis of neutron scattering from lead nuclei. In the simple $\sigma\omega\rho$ model, the Dirac mass is about 0.56. However, modern constraints point to~\cite{Dutra2014,Micaela2017}:

\begin{equation}
    M_N^*/M = 0.6 -0.8 . \label{Diracmass}
\end{equation}

A method to fix both issues simultaneously is to introduce non-linear self-interactions in the $\sigma$ field, as proposed in ref.~\cite{Boguta}:

\begin{equation}
\mathcal{L}_{NL} = - \frac{\kappa M_N (g_{Ns}\sigma)^3}{3} - \frac{\lambda (g_{NS}\sigma)^4}{4}
\end{equation}

The total Lagrangian (in MFA) considering protons, neutrons, electrons, and muons reads:

\begin{eqnarray}
\mathcal{L} = \sum_B\bar{\psi}_B[\gamma^0(i\partial_0 -g_{N\omega}\omega_0 - \frac{1}{2}g_{N\rho}{\tau_3} {\rho}_0) -i\gamma^j \partial_j - M_B^*]\psi_B \nonumber \\-\frac{1}{2}m_s^2\sigma_0^2    + \frac{1}{2}m_\omega^2\omega_0^2
+ \frac{1}{2}m_\rho^2{\rho}_0^2 - \frac{\kappa M_N (g_{Ns}\sigma_0)^3}{3} - \frac{\lambda (g_{NS}\sigma_0)^4}{4}\nonumber \\ +
\sum_l \bar{\psi}_l[(i\gamma^\mu\partial_\mu - m_l)\psi_l \label{NLsigma-omega-rho},
\end{eqnarray}
where the sum in $l$ runs over the leptons ($e,\mu$).

The nucleon eigenvalues, as well as the expected values of the $\omega$ and $\rho$ fields are left unchanged. However, the equation that governs the $\sigma$ field is modified. Applying E-L to Eq.~\ref{NLsigma-omega-rho}:

\begin{equation}
 \sigma_0 = \sum_B \bigg (\frac{g_{Bs}}{m_s^2} \bigg )\langle\bar{\psi}_B\psi_B\rangle  - \bigg (\frac{g_{Ns}}{m_s^2} \bigg ) \bigg [\kappa M_N(g_{Ns}\sigma_0)^2 + {\lambda}(g_{Ns}\sigma_0)^3 \bigg ]
\end{equation}
where $\langle\bar{\psi}_B\psi_B\rangle$ is the traditional scalar density of the baryon $B$ given by Eq.~\ref{scalardensity}.

The reader must be able to show that the total energy density, pressure, and number density are:

 \begin{eqnarray}
  \epsilon = \frac{1}{\pi^2} \sum_B\int_0^{k_{FB}} [\sqrt{M_N^{*2} +k^2}]  k^2 dk \nonumber  + \frac{1}{2}m_s^2\sigma_0^2 + \frac{1}{2}m_\omega^2\omega_0^2  + \frac{1}{2}m_\rho^2\rho_0^2  \nonumber \\ +  \frac{\kappa M_N (g_{Ns}\sigma)^3}{3} + \frac{\lambda (g_{NS}\sigma)^4}{4} + \frac{1}{\pi^2}\sum_l\int_0^{k_{Fl}}\sqrt{m_l^{2} +k^2} k^2 dk ,\nonumber
 \end{eqnarray}

 \begin{eqnarray}
  p = \frac{1}{3\pi^2}\sum_B\int_0^{k_{FB}}\frac{k^4dk}{\sqrt{M_B^{*2} +k^2}}   - \frac{1}{2}m_s^2\sigma_0^2 \nonumber  + \frac{1}{2}m_\omega^2\omega_0^2  
+ \frac{1}{2}m_\rho^2\rho_0^2 \\- \frac{\kappa M_N (g_{Ns}\sigma)^3}{3} -\frac{\lambda (g_{NS}\sigma)^4}{4} +  \frac{1}{3\pi^2}\sum_l\int_0^{k_{Fl}}\frac{k^4dk}{\sqrt{m_l^{2} +k^2}} ,\nonumber
 \end{eqnarray}

 \begin{equation}
   n = \sum_B \frac{k_{FB}^3}{3\pi^2} . \label{eosNLWM}
 \end{equation}

Returning to the matter of compressibility, we recall that $\epsilon/n$ has a minimum at $n =n_0$. Therefore, $K_0$ (Eq.~\ref{Kexp}) represents the concavity of the binding energy per nucleon. A larger value of $K_0$ will produce a larger concavity, resulting in a steeper curve and a stiffer EOS. The inverse is also true. \\

\subsection{Numerical Results}

\begin{figure*}[h!]
\begin{tabular}{ccc}
\centering 
\includegraphics[scale=.58, angle=270]{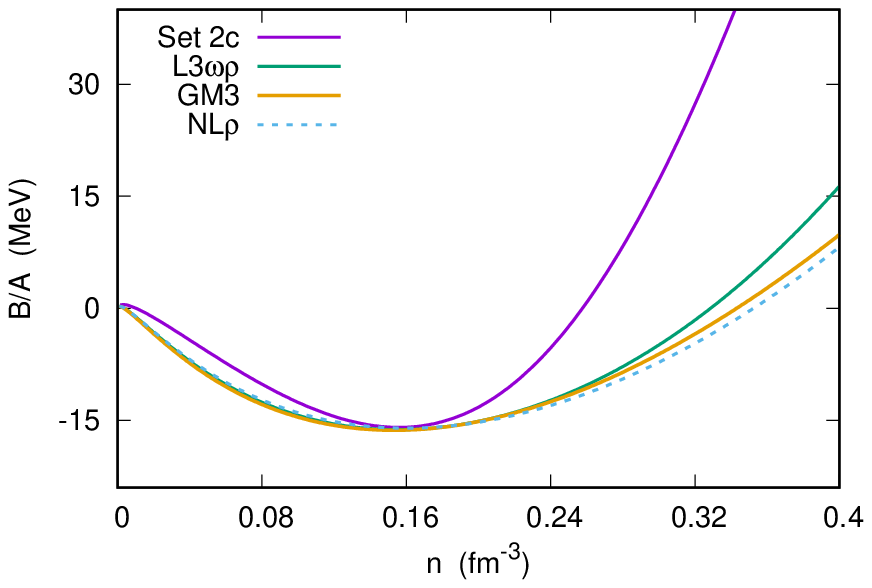} &
\includegraphics[scale=.58, angle=270]{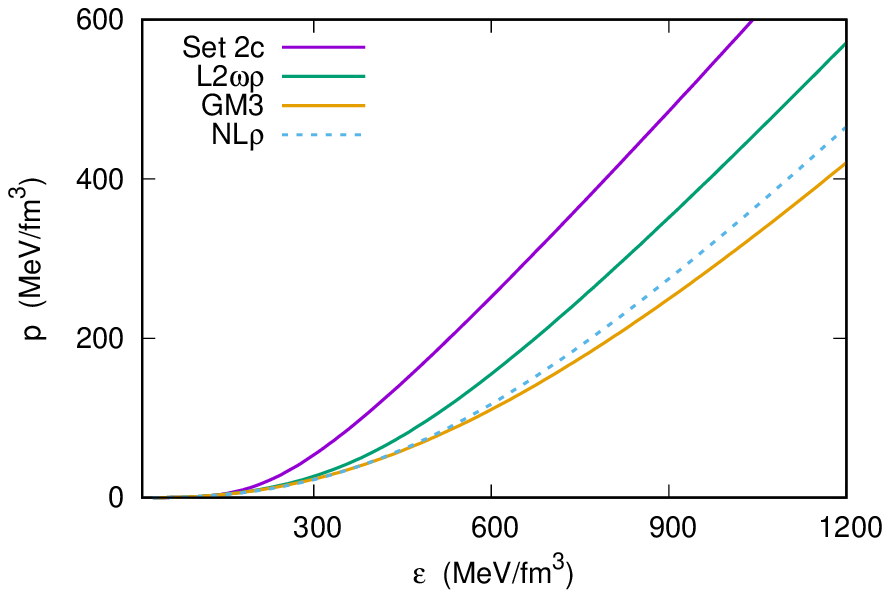} \\
\includegraphics[scale=.58, angle=270]{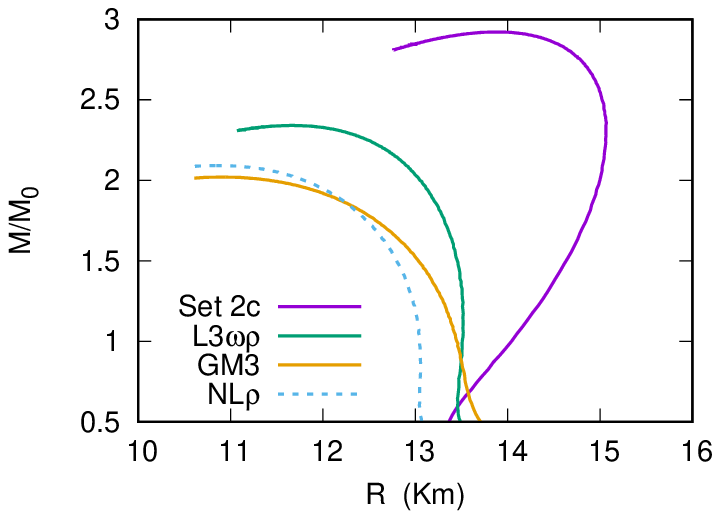} &
\includegraphics[scale=.58, angle=270]{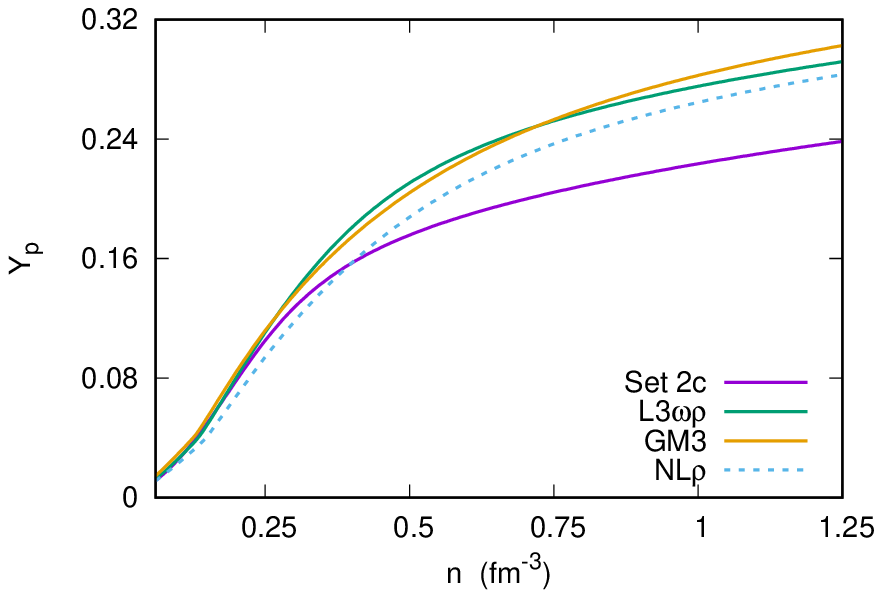} \\
\end{tabular}
\caption{Influence of the $\rho$ field in nuclear matter: (\textbf{a}) Binding energy per nucleon for symmetric matter (dashed lines) and pure neutron matter (solid lines) (\textbf{b}) Proton fractions for different values of $G_\rho$ in beta-stable matter. (\textbf{c}) EOSs. (\textbf{d}) OV solutions.\label{F7}}
\end{figure*}

The non-linear $\sigma\omega\rho$ model of the QHD was a standard approach to relativistic models of neutron stars from the early 1980s~\cite{GLENDENNING1982}, up to the 2020s~\cite{lopesnpa}.  Such models are sometimes referred as non-linear Walelcka models (NLWM)~\cite{Glenbook}~\footnote{Another name is QHD-II, found in some texts~\cite{FabioQHDII}}. We now analyze some well-known parametrizations of the NLWM from a nuclear and stellar point of view. The models, their parameters, and their predictions related to the discussed constraints are presented in Tab.~\ref{T5}, while the numerical results are displayed in Fig.~\ref{F7}.

\begin{center}
\begin{table}[h]
\begin{center}
\begin{tabular}{ccccc|c}
\toprule
\textbf{Physical Quantities}	& Set 2c	& L3$\omega\rho$~\cite{Lopes2022CTP,Lopes2024ApJ}     & NL$\rho$~\cite{Liu2002}  & GM3~\cite{GlenPRL} & Constraints~~~\cite{Dutra2014,Micaela2017}  \\
$- B/A~$ (MeV)      & 15.9     & 16.2  &16.0  &16.3  & 15.8- 16.5 \\     
$n_0~$ (fm$^{-3}$) 	& 0.156		& 0.156	 & 0.160 & 0.153 & 0.148 - 0.170\\
$K_0~$ (MeV)         & 551	    & 256	 & 240   & 240   & 220 - 260 \\
$M^{*}_N/M_N$      & 0.56		& 0.69   & 0.75  & 0.78  & 0.6 - 0.8 \\
$S_0~$ (MeV)       & 32.8       & 31.7   & 30.5  & 32.5   & 30 - 35     \\
\hline
 $M_N~$ (MeV)       & 938.93    &938.93  & 938.93 & 938.93 \\
  $m_s~$ (MeV)       & 512    &512  & 512 & 512 \\
  $m_\omega~$ (MeV)       & 783    &783  & 783 & 783 \\
  $m_\rho~$ (MeV)       & 770    &770  & 770 & 770 \\
    $m_e~$ (MeV)       & 0.51    &0.51  & 0.51 & 0.51 \\
  $m_\mu~$ (MeV)       & 105.66    &105.66  & 105.66 & 105.66 \\
 $G_S~$ (fm$^2$)   &15.00      &12.108 & 10.330 & 9.927 \\
  $G_V~$ (fm$^2$)   &11.40     &7.132 & 5.421 & 4.820 \\
  $G_\rho~$ (fm$^2$)   &3.38    &4.06 & 3.83 & 4.791 \\
  $\kappa~$    & -    & 0.00414& 0.00694 & 0.00866 \\
    $\lambda$   & -    &-0.00390 & -0.00480 & -0.00242 \\

\toprule
\end{tabular}
\caption{Parameters of the models, physical quantities calculated at the saturation point, and constraints found in literature.} \label{T5}
\end{center}
\end{table}
\end{center}

In Fig.\ref{F7} {\bf (a)}, we show the binding energy per nucleon in order to analyze the incompressibility $K_0$. As pointed out earlier, the incompressibility acts as a measure of the "spring constant" in the GMR. The stepper curve produced by Set2c reflects the high value of $K_0$, as such a model does not present non-linear terms in the $\sigma$ field. An interesting feature here is that GM3~\cite{GlenPRL} and NL$\rho$~\cite{Liu2002} have the same value $K_0$ = 240 MeV, but the curve is a little steeper for GM3.

The fact that the curve of $B/A$ is slightly steeper in GM3 than in NL$\rho$ could indicate that GM3 produces an EOS a little stiffer than NL$\rho$. However, this is not true or at least not entirely true. The NL$\rho$ is stiffer at high densities, but the GM3 presents a stiffer EOS until $\epsilon$ = 300 MeV/fm$^{-3}$ as can be seen in {\bf (b)}. The stiffer EOS is Set2c, followed by L3$\omega\rho$~\cite{Lopes2022CTP}. As a general rule, the stiffness of the EOS is determined by the coupling $G_V$. The higher the $G_V$, the stiffer the EOS at high densities (vector meson dominance, or $\omega$ meson dominance).

The fact that NL$\rho$ is stiffer than GM3 at higher densities but softer at low densities suggests that the stars produced by NL$\rho$ are more massive, but with lower radii when compared with GM3. From the OV solutions presented in {\bf (c)}, this is exactly what happens. The maximum mass for the NL$\rho$ is 2.11 $M_\odot$, while it reaches 2.04 $M_\odot$ for the GM3. However, the radius of the canonical 1.4 $M_\odot$ star is 12.92 km for the NL$\rho$ and 13.16 km for the GM3.
L3$\omega\rho$ has a maximum mass of 2.34 $M_\odot$ and a radius of 13.48 km for the canonical star. The ultra-stiffer Set2c has a maximum mass of 2.94 $M_\odot$ with $R_{1.4}$ = 14.52 km. Of course, due to the extreme value of $K_0$, Set2c cannot be perceived as a realistic EOS.

Finally, in {\bf (d)} we display the proton fraction for different models. It can be a little surprising that the Set2c actually predicts the lower amount of protons despite having the higher value of $S_0$. However, with a quick look at Eq.~\ref{Snrho} it is clear that the symmetry energy grows with $G_\rho$. Models with larger values of $G_\rho$ have a larger value of $S(n)$ at high densities and, consequently, a larger proton fraction. The reader is invited to investigate the electron and muon fractions.

Analogous to the $\omega$ dominance, which determines the stiffness of the EOS, we have a $\rho$ dominance for the proton fractions. Higher values of $G_\rho$ will produce a higher value of $Y_p$. \\

 Another nuclear constraint is related to the pressure of symmetric matter at supra-nuclear densities. In ref.~\cite{Daniel2002}, the pressure from 2 to 4.6 times the saturation density was constrained by HIC analyses. On the other hand, ref.~\cite{Steiner2013ApJL} constrained the pressure for densities up to 1.0 fm$^{-3}$ by analyzing transiently accreting neutron stars in quiescence.  
The main problem is that, for high densities, the region constrained in ref.~\cite{Steiner2013ApJL}is
broader than the region constrained in ref.~\cite{Daniel2002} To overcome this
issue, ref.~\cite{Dutra2014} assumes the region from ref.~\cite{Daniel2002} plus an increase of
20$\%$. As presented in ref.~\cite{Dutra2014}, the GM3 and the NL$\rho$ satisfy this constraint in its whole extension. On the other hand, in ref.~\cite{Lopes2022CTP}, it was shown that the L3$\omega\rho$ from 2 to 4.20 times the saturation density, which turns out to be 85$\%$ of the
range. This constraint, nevertheless, has faced some criticism.

\section{Refinements and Astrophysical constraints}

In the early 2000s, new features of the atomic nucleus had arisen. One of them is the so-called neutron skin thickness ($\theta$), which is the difference between the mean value of the neutron radius and the proton radius inside a nucleus~\cite{debora-universe}.

\begin{equation}
\theta = R_n - R_p,
\end{equation}
where $R_n$ and $R_p$ are the mean values of the neutron and proton radii for a given nucleus.
In ref.~\cite{Horo2001}, the authors show that it was possible to control the value of the neutron skin thickness without changing the symmetry energy ($S_0$) by introducing a new non-linear interaction on the Lagrangian. They propose a channel that couples the $\omega$ and the $\rho$ meson:

\begin{equation}
 \mathcal{L}_{\omega\rho} =  \Lambda_{\omega\rho}(g_{N\rho}^2\vec{\rho}^\mu \cdot\vec{\rho}_\mu)\cdot(g_{N\omega}^2\omega^\mu\omega_\mu)   
\end{equation}

Adding this new coupling, the total Lagrangian, in MFA reads~\cite{IUFSU,Miyatsu2013}:

\begin{eqnarray}
\mathcal{L} = \sum_B\bar{\psi}_B[\gamma^0(i\partial_0 -g_{N\omega}\omega_0 - \frac{1}{2}g_{N\rho}{\tau_3} {\rho}_0) -i\gamma^j \partial_j - M_B^*]\psi_B \nonumber \\-\frac{1}{2}m_s^2\sigma_0^2    + \frac{1}{2}m_\omega^2\omega_0^2
+ \frac{1}{2}m_\rho^2{\rho}_0^2 - \frac{\kappa M_N (g_{Ns}\sigma_0)^3}{3} - \frac{\lambda (g_{NS}\sigma_0)^4}{4}\nonumber \\ + \Lambda_{\omega\rho}(g_{N\rho}^2g_{N\omega}^2{\rho_0^2}\omega_0^2)  +
\sum_l \bar{\psi}_l[(i\gamma^\mu\partial_\mu - m_l)\psi_l \label{iufsuL},
\end{eqnarray}

Applying E-L equations, we can obtain the energy eigenvalue for baryons and leptons:

\begin{equation}
  E_B = \mu_B = \sqrt{M_B^{*2} +k_{FB}} + g_{B\omega}\omega_0 + \frac{1}{2}g_{B\rho}\tau_{3B}\rho_0 \label{beev} ,
\end{equation}

\begin{equation}
  E_L = \mu_l = \sqrt{m_l^2 + k_{Fl}^2}  , \label{leev}
\end{equation}

With the help of Statistical mechanics for the fermions and taking $\langle\mathcal{H}\rangle = -\langle \mathcal{L} \rangle$ for the mesons, we obtain the total EOS:

 \begin{eqnarray}
  \epsilon = \frac{1}{\pi^2} \sum_B\int_0^{k_{FB}} [\sqrt{M_N^{*2} +k^2} + g_{N\omega}\omega_0 +\frac{1}{2}g_{N\rho}\tau_3\rho_0]  k^2 dk \nonumber \\  + \frac{1}{2}m_s^2\sigma_0^2 - \frac{1}{2}m_\omega^2\omega_0^2 - \frac{1}{2}m_\rho^2\rho_0^2 \nonumber - \Lambda_v{\rho_0^2}\omega_0^2  \\ +  \frac{\kappa M_N (g_{Ns}\sigma)^3}{3} + \frac{\lambda (g_{NS}\sigma)^4}{4}   \frac{1}{\pi^2}\int_0^{k_{Fe}}\sqrt{m_e^{2} +k^2} k^2 dk. \label{iufsuen}
 \end{eqnarray}
where we define: $\Lambda_v~\equiv~ \Lambda_{\omega\rho}g_{N\omega}^2g_{N\rho}^2$. The expected values for the mesonic fields are obtained by imposing that the energy density is stationary:

\begin{equation} 
\bigg (\frac{\partial \epsilon}{\partial \sigma_0} \bigg ) =
\bigg (\frac{\partial \epsilon}{\partial \omega_0} \bigg ) =
\bigg (\frac{\partial \epsilon}{\partial \rho_0} \bigg ) = 0 , \label{estationarhy}
\end{equation}

\begin{equation}
 \sigma_0 = \sum_B \bigg (\frac{g_{Bs}}{m_s^2} \bigg )\langle\bar{\psi}_B\psi_B\rangle  - \bigg (\frac{g_{Ns}}{m_s^2} \bigg ) \bigg [\kappa M_N(g_{Ns}\sigma_0)^2 + {\lambda}(g_{Ns}\sigma_0)^3 \bigg ]\label{snlinear}
\end{equation}

\begin{equation}
 (m_\omega^2 +2\Lambda_v\rho_0^2)\omega_0 =  \sum_B g_{B\omega}n_B  \label{nlomega} 
\end{equation}

\begin{equation}
 (m_\rho^2 +2\Lambda_v\omega_0^2)\rho_0 =  \sum_B g_{B\rho}\frac{\tau_{3B}}{2}n_B  \label{nlrho} 
\end{equation}

We can combine Eqs.~\ref{nlomega} and \ref{nlrho} with Eq.~\ref{iufsuen} to obtain:

 \begin{eqnarray}
  \epsilon = \frac{1}{\pi^2} \sum_B\int_0^{k_{FB}} [\sqrt{M_N^{*2} +k^2}]  k^2 dk \nonumber  + \frac{1}{2}m_s^2\sigma_0^2 + \frac{1}{2}m_\omega^2\omega_0^2  + \frac{1}{2}m_\rho^2\rho_0^2  \nonumber \\ +  \frac{\kappa M_N (g_{Ns}\sigma)^3}{3} + \frac{\lambda (g_{NS}\sigma)^4}{4} +3\Lambda_v\omega_0^2\rho_0^2\nonumber \\ + \frac{1}{\pi^2}\sum_l\int_0^{k_{Fl}}\sqrt{m_l^{2} +k^2} k^2 dk ,\label{iufsued2}
 \end{eqnarray}
and the pressure is obtained via Eq.~\ref{thermopressure} for the fermions and $p = \langle\mathcal{L}\rangle$ for the mesons:

 \begin{eqnarray}
  p = \frac{1}{3\pi^2}\sum_B\int_0^{k_{FB}}\frac{k^4dk}{\sqrt{M_B^{*2} +k^2}}   - \frac{1}{2}m_s^2\sigma_0^2 \nonumber  + \frac{1}{2}m_\omega^2\omega_0^2  
+ \frac{1}{2}m_\rho^2\rho_0^2 \\- \frac{\kappa M_N (g_{Ns}\sigma)^3}{3} -\frac{\lambda (g_{NS}\sigma)^4}{4} + \Lambda_v\omega_0^2\rho_0^2  \nonumber \\+  \frac{1}{3\pi^2}\sum_l\int_0^{k_{Fl}}\frac{k^4dk}{\sqrt{m_l^{2} +k^2}} .\label{iufsup}
 \end{eqnarray}

 Alternatively, we reinforce that Eq.~\ref{pressure1} is still valid, once we sum over all fermions, $f$, ($npe\mu$):

\begin{equation}
p = \sum_f\mu_f n_f - \epsilon  . \label{pressureAF} .
\end{equation}

We can now analyze the effects of the non-linear $\omega-\rho$ coupling. Besides explicitly contributing to the EOS, the main effect is to change the behavior of the $\omega$ and $\rho$ fields, but in a crossed way. The $\rho$ field affects the $\omega$ field and vice versa. From  Eqs.~\ref{nlomega} and \ref{nlrho}  we can define an effective mass for the $\omega$ and $\rho$ mesons:

\begin{equation}
  m_\omega^{*2} = (m_\omega^2 + 2\Lambda_v\rho_0^2), \quad \mbox{and} \quad   m_\rho^{*2} = (m_\rho^2 + 2\Lambda_v\omega_0^2).
\end{equation}

As the $\omega$ ($\rho$) field increases, it increases the effective mass of $\rho$ ($\omega$) meson and consequently reducing the value of $G_\rho$ ($G_V$). The expressions for the effective mass for the $\omega$ and $\rho$ mesons are similar, but quantitatively they are very different due to the $\omega$ dominance~\footnote{Remembering, that the $\omega$ field is proportional to the total density $n = n_p +n_p$, while the $\rho$ field is proportional to $(n_p -n_n)/2$.}. Consequently, we have a strong reduction of the $\rho$ field but only a gentle change in the $\omega$ one. The reader is invited to numerically display the $\rho$ and $\omega$ fields as a function of the density.

The symmetry energy calculated via Eq.~\ref{Sn2} give us:

\begin{equation}
  S(n) = \frac{n}{8} \bigg (\frac{g_{N\rho}}{m_\rho^*} \bigg )^2  + \frac{k_F^2}{6\sqrt{M^{*2} + k_F^2}}. \label{SnrhoNL}
\end{equation}

Although similar to Eq.~\ref{Snrho}, the increase of the $m_\rho^*$ causes the symmetry energy $S(n)$ to present a much gentler variation. Now, if the symmetry energy at the saturation point is kept fixed, what parameter varies as we vary $\Lambda_{\omega\rho}$? The answer is the symmetry energy slope, $L$, given by~\cite{Rafa2011,lopescesar,Roca2011}:

\begin{equation}
 L = 3n \bigg ( \frac{dS}{dn} \bigg ) \bigg |_{n =n_0} .   \label{slope}
\end{equation}

 The neutron skin thickness is a complex quantity, and the theoretical prediction of its value is ultimately model-dependent. Nevertheless, it is strongly correlated to the slope of the symmetry energy. The larger the value of $L$, the larger is the neutron skin thickness.
Unlike the other five parameters at the saturation point ($M^*_N/M_N,~S_0~,B/A~,K_0$, and $n_0$), $L$ is poorly constrained nowadays. Our ignorance about the true value of $L$ is well expressed in ref.~\cite{Tagami2022}. While the CREX group predicts 0 MeV$ ~<L~<51$ MeV by analyzing the neutron skin thickness of $^{48}$Ca, the PREX group predicts 65 MeV$ ~<L~<165$ MeV by analyzing the neutron skin thickness of $^{208}$Pb. The results do not even overlap.

Nevertheless, $L$ was strongly constrained by combining astrophysical data with nuclear properties measured in the PREX-II experiment, together with chiral effective field theory in ref.~\cite{Essick2021}:

\begin{equation}
    38~\mbox{MeV}~<~L~<~67~ \mbox{MeV}. ~\label{Lcons}
\end{equation}

We use this range as a constraint, but we also present results with higher values of $L$ to better understand its effects.\\

\subsection{{Astrophysical Constraints}}

Up to the mid-2000s, constraints related to neutron stars' observations were virtually nonexistent. Our most reliable bound was the mass of the Vela X-1 pulsar, $M = 1.88~\pm0.13~M_\odot$~\cite{VelaX1}. This indicated that as long as the constraints related to nuclear physics were fulfilled, any EOS that predicts a maximum mass above 1.65 $M_\odot$ could be considered realistic. Not a hard task.

The situation began to change in early 2010, with the discoveries of the PSR J1614-2230 with a mass of 1.97 $\pm$ 0.04 $M_\odot$~\cite{Demorest2010} and the PSR J0348+0432  with a mass of 2.01 $\pm$ 0.04 $M_\odot$~\cite{Antoniadis}. These pulsars provide unequivocal proof that two-solar-mass pulsars exist. However, the real breakthrough came with the launch of the NICER (Neutron star Interior Composition ExploreR) X-ray telescope~\cite{NICER2012}. Two NICER teams~\cite{Miller2021,Riley2021} measured not only the mass but also the radius of the PSR J0740+6620, indicating that they lie in the range 2.08 $\pm$ 0.07 $M_\odot$ and 11.41 km < $R$ < 13.69 km respectively. The possible existence of even more massive pulsars is still under debate~\cite{romani,Lopes2022ApJ}.

Another strong constraint coming from the NICER observations is the radius of the canonical 1.4 $M_\odot$ neutron stars. In ref.~\cite{Miller2021}, the authors point out that $R_{1.4} = 12.45~\pm~0.65$ km, which limits the radius of the canonical star within an uncertainty of only 5\%.

A third constraint is related to the cooling of the neutron stars. The standard model of neutron star cooling is based upon neutrino emission from the interior that is dominated by the modified URCA process:

\begin{equation}
  n + n \to n +p +e +\bar{\nu}_e.  
\end{equation}

However, neutron stars could cool way faster if the direct URCA process occurs in the neutron stars' interiors.

\begin{eqnarray}
    n \to p + e +\bar{\nu}_e.
\end{eqnarray}

It was shown in ref.~\cite{Lattimer1991} that the direct URCA (DU) process can be enabled in neutron stars' interior if the proton fraction exceeds
some critical value ($x_{DU}$) around 11\% to 15\%~\cite{Dohi2019,lopes2024PRCb}: 

\begin{equation}
    x_{DU} = \frac{1}{1+(1 +x_e^{1/3})^3} , \label{xdu}
\end{equation}
where $x_e = n_e/(n_e + n_\mu)$, and $n_e$, $n_\mu$ are the number densities of the electron and muon, respectively.

Reference~\cite{klahn2006} pointed out that any acceptable
EoS does not allow the direct URCA process to occur in neutron stars with masses below 1.5 $M_\odot$. We use this assertion $(M_{DU}~>1.5M_\odot)$ as the third constraint.\\

\subsection{Numerical Results}

\begin{figure*}[h!]
\begin{tabular}{ccc}
\centering 
\includegraphics[scale=.58, angle=270]{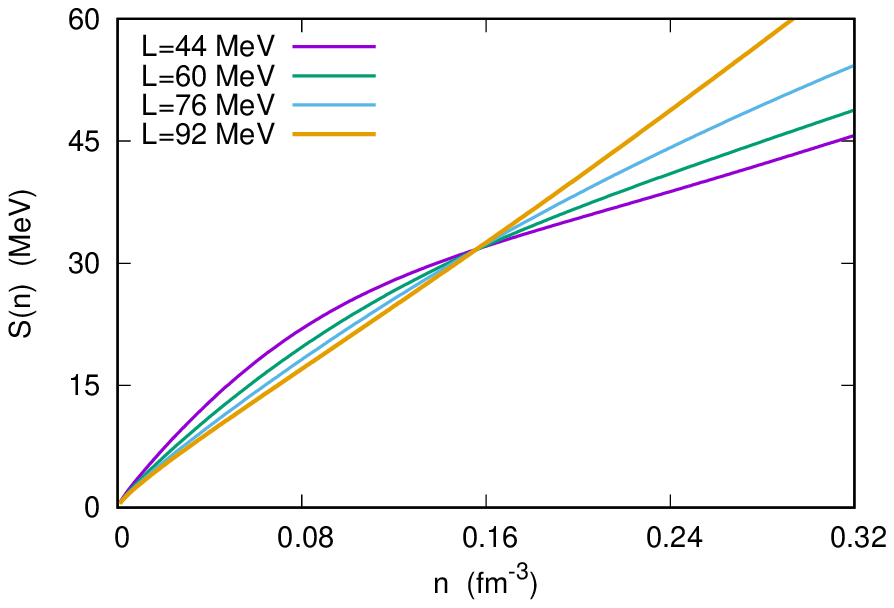} &
\includegraphics[scale=.58, angle=270]{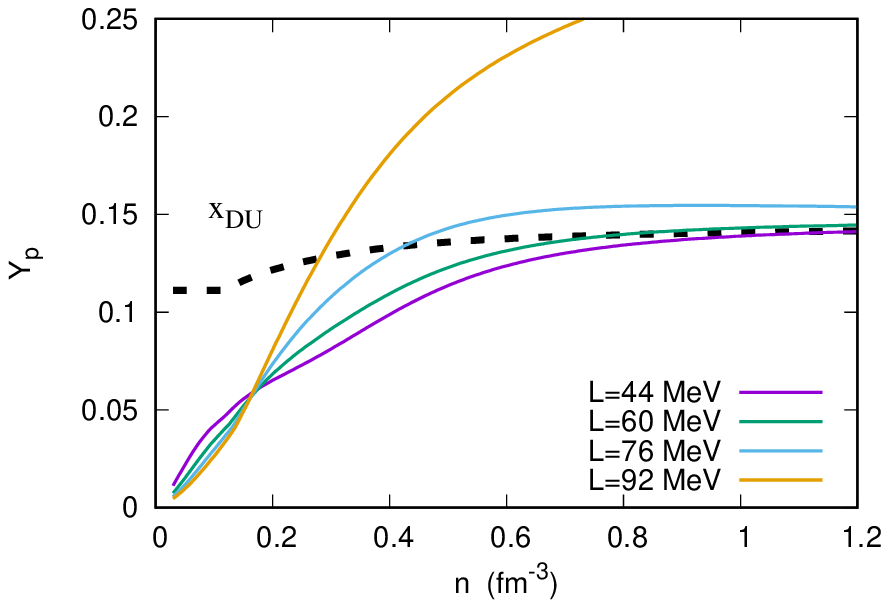} \\
\includegraphics[scale=.58, angle=270]{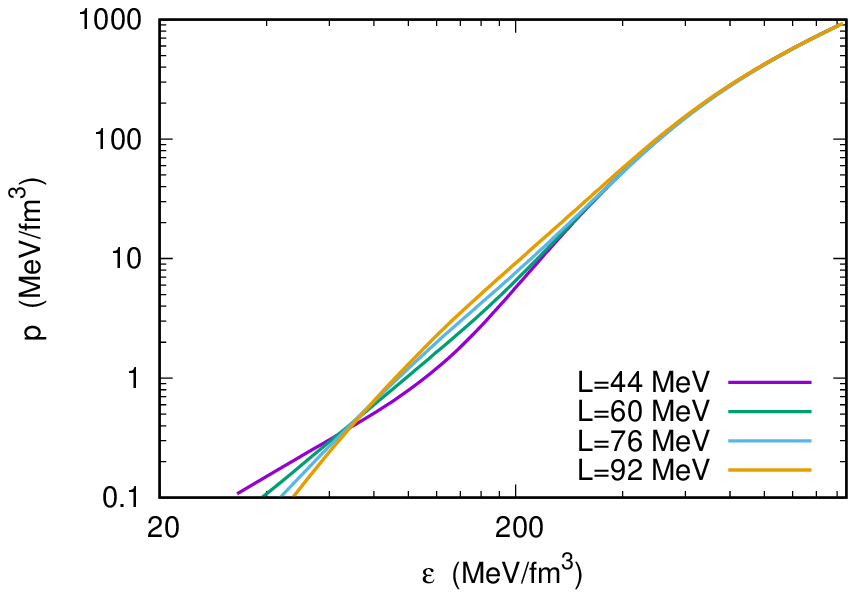} &
\includegraphics[scale=.58, angle=270]{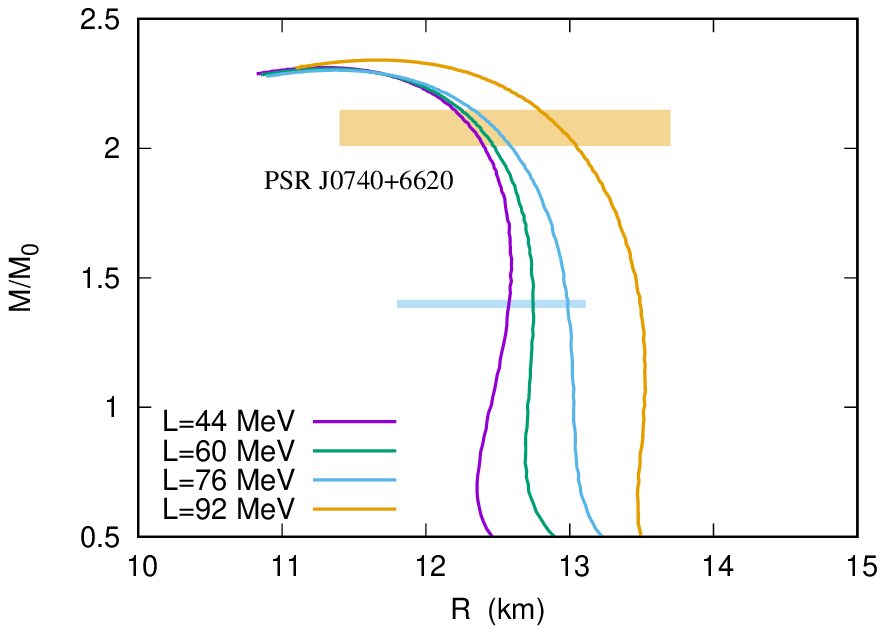} \\
\end{tabular}
\caption{Influence of the slope in nuclear matter: (\textbf{a}) Symmetry energy $S(n)$. (\textbf{b}) Proton fraction in beta-stable matter. (\textbf{c}) EOSs. (\textbf{d}) OV solutions and astrophysical constraints related to the PSR J0740+6620 and the canonical star.\label{F8}}
\end{figure*}

We now analyze the effects of $\Lambda_{\omega\rho}$ in symmetric and beta-stable nuclear matter. We use only the L3$\omega\rho$ model~\footnote{Actually, the original L3$\omega\rho$ presented in ref.~\cite{Lopes2022CTP} already has the $\Lambda_{\omega\rho}$ term.}, as pointed out in ref.~\cite{lopes2024PRCb}, changing the model does not produce qualitative differences. 
Furthermore, all five physical quantities presented in Tab.~\ref{T5} ($M^*_N/M_N,~S_0~,B/A~,K_0$, and $n_0$) are the same here. Only the slope ($L$) is varied. To accomplish this task, all the constants presented in Tab.~\ref{T5} are also the same. The only exception is $G_\rho$ that must be modified in order to keep $S_0$ = 31.7 MeV, as we vary $\Lambda_{\omega\rho}$. The numerical results are presented in Fig.~\ref{F8}, while the values of $G_\rho$ and $\Lambda_{\omega\rho}$ and the main properties of neutron stars are summarized in Tab.~\ref{T6}.

\begin{center}
\begin{table}[h]
\begin{center}
\begin{tabular}{ccccccc}
\toprule
\			{$L~$} (MeV)	& $G_\rho$ (fm$^2$)  & $\Lambda_{\omega\rho}$ & $n_{DU}~$ (fm$^{-3}$)    & $M_{max}/M_\odot$  & $R_{1.4}~$ (km) & $M_{DU}/M_\odot$  \\
 44      & 8.40    &  0.0515  &1.27  &2.31  & 12.58 & -\\     
60	& 6.16		&  0.0344	 & 0.80 & 2.30 & 12.74 & 2.28\\
76    & 4.90	    & 0.0171 & 0.44  & 2.30   & 12.99 & 1.62 \\
92     &  4.06		& 0.0   & 0.29  & 2.34  & 13.48 & 1.10 \\
\toprule
\end{tabular}
\caption{Different values of the slope and the parameters $G_\rho$ and $\Lambda_{\omega\rho}$, as well some neutron stars' properties within the L3$\omega\rho$ model}\label{T6}
\end{center}
\end{table}
\end{center}

In Fig~\ref{F8} {\bf (a)} we display the symmetry energy $S(n)$ according to Eq.~\ref{SnrhoNL}. As can be seen, the value of the $S_0$ is the same for all parametrizations, once we impose it. For densities below the saturation point, we see that the symmetry energy is higher for low values of $L$ and vice versa. 
For densities above the saturation point, the behavior of $S(n)$ is inverted.
This can be explained by Eq.~\ref{SnrhoNL}, along with the values of $G_\rho$ presented in Table~\ref{T6}. As lower values of $L$ require higher values of $\Lambda_{\omega\rho}$ and consequently higher values of $G_\rho$, at low densities, where $m_\rho^* ~\approx~m_\rho$, the first term in Eq.~\ref{SnrhoNL} dominates. As the density increases, the effective mass of the $\rho$ meson also increases, reducing the contribution of the first term. The higher the value of $\Lambda_{\omega\rho}$, the higher is the reduction of the symmetry energy at high densities.

In {\bf (b)} we present the proton fraction, $Y_p$. As it strongly depends on the symmetry energy, the discussion is analogous to Fig. {\bf (a)}. Lower slope values will have higher values of $Y_p$ for $n~<n_0$ and vice versa.
The new feature here is to investigate the density where the proton fraction reaches $x_{DU}$, which we call $n_{DU}$. For large slope values (or low values of $\Lambda_{\omega\rho}$), $x_{DU}$ is reached at low densities. For example, for $\Lambda_{\omega_\rho}$ = 0, which produce $L = 92$ MeV, we have $n_{DU} = 0.29$ fm$^{-3}$, which is less than twice the saturation density. This produces $M_{DU}$ = 1.10 $M_\odot$, which is in disagreement with the constraint  $M_{DU}~>1.5$ $M_\odot$ discussed in ref.~\cite{klahn2006}.  An interesting point is that for $L = 44$ MeV, $n_{DU} = 1.27$ fm$^{-3}$. This value is above the central density reached in the maximum mass neutron star. Therefore, in this case, the DU process is never enabled.  It is worth pointing out that the curves related to $x_{DU}$ are model dependent, although their variation are small. In {\bf (b)}, the dotted $x_{DU}$ curve is related to $L = 60$ MeV, and it is presented to give us an estimate of the value of $n_{DU}$.

In {\bf (c)} we display the EOSs in log scale in order to recognize the differences easily.
As can be seen, lower values of $L$ have produced stiffer EOSs at low densities, but the situation is reversed in the intermediary regime. As densities increase, the EOSs degenerate due to $\omega$ dominance.  Soft EOSs at intermediate densities followed by degenerate EOSs at
high densities indicate that lower values of $L$ will produce neutron stars with smaller radii but with similar maximum masses to those with high values of $L$.

Finally, in {\bf (d)}, the OV solution is presented along the NICER constraints related to the PSR J0740+6620 and the canonical 1.4$M_\odot$ star. As expected due to the behavior of the EOSs, lower values of $L$, predict small radii for neutron stars but very similar maximum masses, once all models have the same $G_V$ value. The study of the influence of slope on neutron star properties is a hot topic in the modern literature~\cite{Rafa2011,lopescesar,Lopes2024ApJ}.

Concerning the astrophysical constraints, we see that one of the advantages of the L3$\omega\rho$ parametrization is that it can fulfill the constraints related to the  PSR J0740+6620 with several values of $L$. In relation to the canonical star, we can obtain $R_{1.4}~<$ 13.1 km for all values of $L$ but $L$ = 92 MeV. The values of $L = 44$ MeV and $L = 60$ MeV virtually satisfy every constraint in both the nuclear and astrophysical realms. \\

\section{Conclusions} \label{conclusions}

In this work, we introduce the QHD to late undergraduate and early graduate students, focusing on the investigation of neutron stars' interiors. 
We begin with the free Fermi gas and end in a realistic model that virtually satisfies all constraints from nuclear experiments and astrophysical observations. The main results can be summarized as follows:

\begin{itemize}
    \item We present the hydrostatic equilibrium equations for both the Newtonian and full relativistic approaches. There are two equations but three variables ($\epsilon$, p, and $M(r)$). Hence, the need for an EOS.

    \item The formalism to calculate the energy density of fermions via statistical mechanics is introduced, as well as the Dirac equation and Lagrangian. The EOS for free neutrons is presented, producing a maximum mass of 0.71 $M_\odot$.

    \item We introduce the scalar $\sigma$ meson, responsible for the attraction between the nucleons and, therefore, for the stability of the atomic nuclei. Its expected value is calculated in MFA, as well as the EOS. The $\sigma$ field softens the EOS.

    \item The repulsive vector $\omega$ meson is introduced to account for the saturation of the nuclear force, as well the discussion about its effects and how to calculate its expected value. The $\sigma-\omega$ model is discussed. Due to its vector nature, the $\omega$ meson dominates at high densities ($\omega$ dominance).

    \item SEMF is introduced, as well as the first nuclear constraints that the $\sigma-\omega$ model must satisfy ($B/A$ and $n_0$). The neutron star crust and beta-stable matter are discussed. The first minimally realistic stars are built.

    \item The vector-isovector $\rho$ meson is introduced to fix the symmetry energy parameter. Its effects on the proton fraction, EOSs, and neutron stars' macroscopic properties are discussed.

    \item The possible existence of muons in neutron stars' interiors is discussed, as well as how to incorporate them in beta-stable matter. Self-interaction of the $\sigma$ meson is introduced to correct the incompressibility. Standard models of the QHD able to satisfy the five main constraints of the nuclear matter ($M^*_N /M_N~,S_0~,B/A~,K_0,$ and $n_0$) are introduced.

    \item A sixth parameter is introduced, the symmetry energy slope ($L$). We show how to control it via a non-linear coupling between the $\omega$ and $\rho$ mesons. Astrophysical constraints are presented, as well as a model that is able to fulfill virtually all constraints related to nuclear and astrophysical observations.

\end{itemize}

 In the upcoming years, we expect that new observations will lead us to new and more precise bounds in observations of neutron stars. A detailed study of the cooling of neutron stars can lead us to constrain not only the presence or not of DU effect, but also the presence of superfluidity in the neutron stars' core~\cite{Yakovlev2004b,YAKOVLEV2004a}. Another approach to investigate superfluidity is the study of pulsar glitches~\cite{Marco2022}. Finally, the study of the tidal deformation coming from gravitational wave observatories has provided us with strong constraints on the EOSs in the last years~\cite{Chat2020,Hinderer_2008,Flores2020}.  It was inferred that the dimensionless tidal parameter for the canonical star lies  $70~<~\Lambda_{1.4}~<580$~\cite{AbbottPRL,Abbott:2018wiz}. As discussed in ref.~\cite{lopes2024PRCb}, L3$\omega\rho$ with $L$ = 60 MeV has $\Lambda_{1.4} = 513$, fulfilling this constraint.

Before we finish the paper, we point out that the theory of neutron stars is far from over. Due to the extreme density reached in their interior, exotic matter may be present at their core. Hyperons are the main candidate, as some studies point out that their presence is inevitable~\cite{Dapo2010,lopesnpa,WeissPRC2012,Jiang2012}. Other possibilities include $\Delta$-resonances~\cite{lopesPRD,BETHE1974} and kaons condensate~\cite{kaonp}. A more exotic scenario suggests that the ordinary matter as we know it, composed of protons and neutrons, is only metastable.
In this case, the true ground state of matter is the so-called strange matter, composite by deconfined quarks. A supernova explosion could trigger the deconfinement, and the observed pulsars would actually be strange stars, formed by deconfined $uds$ quarks~\cite{Olinto,lopesEPJC2025}.

A detailed discussion about these subjects is left for future works.

\acknowledgments{ L.L.L.  was partially supported by CNPq (Brazil)
under Grant No 305347/2024-1. }

\bibliographystyle{ieeetr}
\bibliography{aref}

\end{document}